\documentclass[prb,reprint,a4paper,citeautoscript,floatfix]{revtex4-2}

\pdfoutput=1						
\usepackage[charter]{mathdesign}	
\usepackage{amsmath}				

\usepackage[pdftex]{graphicx}
\usepackage{microtype}
\usepackage{bm}\let\vec\bm

\usepackage[svgnames]{xcolor}
\usepackage[colorlinks=True,linkcolor=DarkRed,citecolor=ForestGreen,urlcolor=MediumBlue,
	pdfstartview=FitH,bookmarks=False,pdfpagemode=UseNone]{hyperref}

\def\kB{k_{\mathrm{B}}}
\def\RH{R_{\mathrm{H}}}
\def\nH{n_{\mathrm{H}}}

\begin{document}

\title{Universal low-density power laws of the dc conductivity and Hall constant in the self-consistent Born approximation}

\author{Giacomo Morpurgo}
\affiliation{Department of Quantum Matter Physics, University of Geneva, 1205 Geneva, Switzerland}
\author{Christophe Berthod}
\affiliation{Department of Quantum Matter Physics, University of Geneva, 1205 Geneva, Switzerland}
\author{Thierry Giamarchi}
\affiliation{Department of Quantum Matter Physics, University of Geneva, 1205 Geneva, Switzerland}

\date{July 8, 2025}

\begin{abstract}

The dc conductivity tensor of two-dimensional one-band metals with weak pointlike disorder and magnetic field is studied in the self-consistent Born approximation, with special emphasis on the regime of low carrier density. In this theory, the Kubo conductivity is a functional of the electron dispersion and local (momentum-independent) electron self-energy, which is itself a causal functional of the dispersion and disorder strength. We obtain closed expressions for the asymptotic low-density conductivities at zero temperature in the form of power laws of the density and disorder strength with universal exponents. The crossover to the semiclassical regime of high density is studied numerically, as well as the temperature dependence. Our model and results may be relevant to interpret linear magneto-transport experiments performed in the metallic regime of gated two-dimensional semiconductors.

\end{abstract}

\maketitle

\section{Introduction}

After the pioneering works on two-dimensional (2D) organic and conventional semiconductors in the 1980s \cite{Review_1982_Ando_2d_semiconductors, Tsumura_Ando_1986_pionnier_field_effect_transistor}, the experimental research on field-effect doping has accelerated since the discovery of graphene \cite{Novoselov_Firsov_2004_graphene_field_effect}, monolayer transition-metal dichalcogenides \cite{Review_Manzeli_Kis_2017_TMD}, and more recently moir\'{e} systems \cite{Cao_Jarillo_Herrero_2018_bilayer_graphene_magic_angle, Review_Andrei_Young_2021_Moire_materials}, transforming our understanding of 2D materials and enabling new technologies. From a technological viewpoint, two of the most relevant quantities in 2D gated conductors are the carrier density, which is varied by applying the gate voltage, and the carrier mobility, which depends on the intrinsic scattering mechanisms and the sample purity. While these two quantities govern the transport properties of the devices, none of them can be directly measured.

Two different methods are commonly used to estimate the carrier density. The first relies on the Hall constant being inversely proportional to the carrier density, as found in the classical Drude and semiclassical Boltzmann theories, while the second relies on modeling the device capacitance \cite{Kopp_Mannhart_2009_compute_capacitance, Li_Levi_2011_organic_measure_density_mobility}. Both approaches have limitations. The universality of the classical relation between the Hall constant and the carrier density has been repeatedly questioned \cite{Fukuzawa_Ando_2009_Hall_effect_graphene_SCBA, Noro_Ando_2016_Hall_long_range_scattering_graphene, Morpurgo_Giamarchi_2024_Hall_effect_low_density}. On the other hand, modeling the capacitance with geometrical and quantum terms was shown to disregard contributions that are potentially important in ultrathin devices \cite{Zhang_Morpurgo_2019_cross_quantum_capacitance, Berthod_Giamarchi_2021_cross_quantum_capacitance}. Similar concerns have been raised regarding the estimation of the mobility \cite{Choi_Podzorov_2018_measure_charge_carrier_mobility_accurate}, which is meant to be a density-independent measure of the scattering.

To uncover the precise relationship linking carrier density and scattering with the transport characteristics in regimes where the conventional rules are suspected to fail, in particular in the regime of low carrier density \cite{Morpurgo_Giamarchi_2024_Hall_effect_low_density}, one has to study rigorously the conductivity tensor and the Hall effect, which remains a hard problem to this date. The semiclassical transport theory \cite{Book_Ziman_2001_transport_electron_and_phonons} may be justified when the scattering rate is the smallest energy scale, making it unsuitable to study transport at vanishingly low carrier densities, where the smallest energy scale is the Fermi energy. On the other hand, the rigorous microscopic quantum transport theory \cite{Fukuyama_Wada_1969_Hall_effect_Kubo_I, Fukuyama_1969_Hall_effect_Kubo_II, Itoh_1990_Hall_effect_ladder_approximation} remains intractable in the thermodynamic limit, although interesting perturbative \cite{Khodas_Finkelstein_2003_Hall_coefficient_ee_interactions, Kovacevic_Vucicevic_2025_vertex_correction_Hubbard} and nonperturbative \cite{Prelovsek_Zotos_2001_Hall_constant_correlated_systems, Auerbach_2018_Hall_number_metals_via_susceptibilities} results have been reported. For nearly half-filled bands, progress has been possible for Hubbard-like models using exact numerical techniques on small lattices \cite{Castillo_Balseiro_1992_Hall_Hubbard_large_U, Assaad_Imada_1995_Hall_2D_Hubbard, Wang_Deveraux_2020_Hall_effect_Hubbard_QMC} or approximate theories \cite{Pruschke_Freericks_1995_tranport_Hubbard_DMFT, Markov_Rubstov_2019_Hofstadter_butterfly_Hubbard_DMFT, Vucicevic_Zitko_2021_DMFT_conductivity_Hubbard, Manning_Chen_1993_Hall_effect_t_J_model, Kontani_Ueda_1999_Hall_effect_high_Tc, Kontani_2007_Hall_effect_high_Tc_FLEX, Mitscherling_Metzner_2018}, as well as for quasi-one dimensional models \cite{Leon_Giamarchi_2006_Hall_quasi_1d_organic, Leon_Giamarchi_2007_Hall_strong_correlated_low_d_systems, Leon_Millis_2008_Hall_triangular_lattice} and ladders \cite{Filippone_Giamarchi_2019_vanishing_Hall, Greschner_Giamarchi_2019_Universal_Hall_response, Buser_Giamarchi_2021_Hall_voltage_synthetic_quantum_systems}.

In comparison, the magneto-transport in the regime of low carrier density, which is accessible experimentally via field-effect doping of semiconductors or semimetals, has received little theoretical attention so far. This is understandable, as some of the approximations that enable analytical treatments of microscopic Hamiltonians at high density are no longer allowed, while among the numerical techniques, only those working in the thermodynamic limit have a chance to access this regime. An intermediate class of tractable theories, where both analytical and exact numerical results can be obtained at low density, is formed by the locally-correlated models, where the conductivity only depends on the dressed electron Green's function \cite{Khurana_1990_conductivity_infinite_d_Hubbard, Markov_Rubstov_2019_Hofstadter_butterfly_Hubbard_DMFT, Vucicevic_Zitko_2021_DMFT_conductivity_Hubbard}. 

In the present work, we study one such locally correlated model describing impurity scattering. At variance with previous studies dealing with impurity scattering \cite{Liu_1970_Kubo_conductivity_impurity, Fukuyama_1980_Hall_effect_2d_disordered, Fukuzawa_Ando_2009_Hall_effect_graphene_SCBA, Ando_2014_Hall_long_range_scattering_bilayer_graphene, Noro_Ando_2016_Hall_long_range_scattering_graphene}, we define here a model that is meant to be local from the outset---by using pointlike impurities treated in the self-consistent Born approximation---and we solve it without further approximation. We derive asymptotic results at low density and zero temperature, which we back up by accurate numerical calculations. Both for parabolic and cosine dispersions in two dimensions, we find that the conductivity and Hall constants are power laws of the density with universal exponents, $\sigma_{xx}\sim n^{2/3}$ and $\RH\sim n^{-1/3}$, whereas the usual laws $\sigma_{xx}\sim n$ and $\RH\sim n^{-1}$ are recovered at high density. Assuming that the change in gate voltage translates into the same change in chemical potential, a contact is established with the field-effect experiments. We show that the chemical potential follows a power law as well, $\mu-\varepsilon_0\sim n^{2/3}$ with $\varepsilon_0$ a model-dependent threshold, which implies a linear relation $\sigma_{xx}\sim\mu-\varepsilon_0$ at low density. We finally study the temperature dependence and find it mostly featureless and strongest at low density.

The paper is organized as follows. Section~\ref{sec:theory} reviews the basic equations that define the linear conductivity tensor in a one-band model of locally correlated electrons (\ref{sec:bubble}), discusses the drawbacks of the phenomenological scattering model characterized by a constant scattering rate (\ref{sec:Gamma}), and describes the microscopic scattering model used in the present work (\ref{sec:SCBA}). Section~\ref{sec:results} reports our numerical and analytical results for $\sigma_{xx}$, $\sigma_{xy}$, and $\RH$ versus density and coupling constant for two representative electron dispersion relations in two dimensions. A discussion of the relevance of our model and results to interpret real field-effect experiments is provided in Sec.~\ref{sec:discussion}, followed by conclusions and perspectives in Sec.~\ref{sec:conclusion}. Additional details about the numerical method and the scattering model can be found in Appendixes~\ref{app:piecewise}, \ref{app:ED}, and \ref{app:edge}.

\section{Models and methods}
\label{sec:theory}

We evaluate the longitudinal and transverse electrical conductivities for two-dimensional metals in a perpendicular magnetic field, which allows us to deduce the Hall constant, using the Kubo formula and describing the scattering with a causal (i.e., Kramers-Kronig consistent) momentum-independent self-energy. Within this framework, the Kubo conductivity tensor can be evaluated exactly up to first order in the magnetic field, using expressions that take the noninteracting electron dispersion $E_{\vec{k}}$ and the self-energy $\Sigma(\varepsilon)$ as the only system-dependent inputs \cite{Fukuyama_Wada_1969_Hall_effect_Kubo_I,Fukuyama_1969_Hall_effect_Kubo_II, Morpurgo_Giamarchi_2024_Hall_effect_low_density, Pickem_Tomczak_2022}. 

In this work, we consider two representative dispersions: A parabolic dispersion describing nearly free electrons or Bloch electrons sufficiently close to a conduction-band minimum, as well as a square-lattice cosine dispersion, which also becomes parabolic near the band edge. In each of these two cases, the scattering is supposed to originate from a distribution of identical pointlike impurities and it is treated by means of the impurity-averaging technique within the self-consistent Born approximation (SCBA) \cite{Book_Bruus_Flensberg_2016_Many_Body, Book_Berthod_2018_Spectroscopic_Probes}. 

We start this section by recalling the relevant general expressions that provide the conductivities and we briefly explain the numerical methods that we use to compute them accurately and efficiently. We then show the limitations of the phenomenological self-energy model $\Sigma(\varepsilon)=-i\Gamma$ in the low-density regime. In a last step, we define the SCBA self-energy and we display its properties, exhibiting the similarities and the differences that occur for parabolic and cosine dispersions.

\subsection{Kubo conductivities and Hall constant for locally correlated two-dimensional metals}
\label{sec:bubble}

The dependence of the conductivity tensor on the noninteracting electron dispersion $E_{\vec{k}}$ is captured by three functions of the energy $E$. The first is the noninteracting density of states (DOS), which counts the number of states per unit surface at energy $E$, using the Dirac $\delta$ function:
	\begin{equation}\label{eq:N0}
		N_0(E)=2\int\frac{d^2k}{(2\pi)^2}\,\delta\left(E-E_{\vec{k}}\right).
	\end{equation}
The factor $2$ accounts for the electron spin. The second function is the longitudinal transport function, which also counts states, however weighted by the square of their group velocity in direction $x$ multiplied by the electron charge $e$:
    \begin{equation}\label{eq:Phi0}
		\Phi_{xx}(E)=\left(\frac{e}{\hbar}\right)^22\int\frac{d^2k}{(2\pi)^2}
		\left(\frac{\partial E_{\vec{k}}}{\partial k_x}\right)^2\delta(E-E_{\vec{k}}).
	\end{equation}
The third function is the transverse transport function, where now states are weighted by a measure of the dispersion curvature:
    \begin{multline}\label{eq:Phi1}
		\Phi_{xy}(E)=\left(\frac{|e|}{\hbar}\right)^3\frac{2\pi^2}{3}\int\frac{d^2k}{(2\pi)^2}
		\left[2\frac{\partial E_{\vec{k}}}{\partial k_x}
		\frac{\partial E_{\vec{k}}}{\partial k_y}
		\frac{\partial^2E_{\vec{k}}}{\partial k_x\partial k_y}\right.\\
		\left.-\left(\frac{\partial E_{\vec{k}}}{\partial k_x}\right)^2
		\frac{\partial^2E_{\vec{k}}}{\partial k_y^2}
		-\left(\frac{\partial E_{\vec{k}}}{\partial k_y}\right)^2
		\frac{\partial^2E_{\vec{k}}}{\partial k_x^2}\right]\delta(E-E_{\vec{k}}).
	\end{multline}
These three functions are known analytically for a parabolic band, but for more general dispersions, like in the cosine case, they must be evaluated numerically. As they depend neither on the scattering mechanism, nor on the electron density or the temperature at which the conductivities are being evaluated, they only need to be calculated once when scanning these parameters. We have found it to be convenient and numerically efficient to represent these functions as piecewise-continuous functions that are fast to evaluate at any energy $E$ \cite{Berthod_2025_Piecewise}. In each piece, well-chosen mathematical functions are optimized and reproduce with more than six-digit fidelity the DOS and transport functions computed numerically by quadratures. Appendix~\ref{app:piecewise} provides an illustration for $N_0(E)$.

The conductivities depend on the complex self-energy $\Sigma(\varepsilon)$ through the electron spectral function
    \begin{equation}\label{eq:A}
		A(E,\varepsilon)=\frac{-\mathrm{Im}\,\Sigma(\varepsilon)/\pi}
    	{[\varepsilon-E-\mathrm{Re}\,\Sigma(\varepsilon)]^2+[\mathrm{Im}\,\Sigma(\varepsilon)]^2},
	\end{equation}
which describes the redistribution of electronic spectral weight by interactions. The spectral weight is entirely concentrated at $\varepsilon=E$ in the absence of scattering [$\Sigma(\varepsilon)=-i0$], while otherwise it is spread around the quasiparticle energy $E^*=E+\mathrm{Re}\,\Sigma(E^*)$ over an energy range typically given by $-\mathrm{Im}\,\Sigma(E^*)$. The noninteracting DOS and the spectral function allow one to compute the interacting DOS,
	\begin{equation}\label{eq:N}
		N(\varepsilon)=\int_{-\infty}^{\infty}dE\,N_0(E)A(E,\varepsilon),
	\end{equation}
which differs from $N_0(\varepsilon)$ by corrections scaling with the amplitude of the self-energy. Calculations performed at fixed carrier density $n$ require us to determine the chemical potential $\mu$ according to
    \begin{equation}\label{eq:n}
		n=\int_{-\infty}^{\infty}d\varepsilon\,f(\varepsilon-\mu)N(\varepsilon),
	\end{equation}
where $f(\varepsilon)=(e^{\varepsilon/\kB T}+1)^{-1}$ is the Fermi-Dirac distribution for temperature $T$ and $\kB$ is the Boltzmann constant.

With the chemical potential set, the longitudinal and transverse conductivities for a magnetic field $B$ follow as
	\begin{align}
		\label{eq:sigmaxx}\hspace{-0.7em}
    	\sigma_{xx}&=\pi\hbar\int_{-\infty}^{\infty}d\varepsilon\,[-f'(\varepsilon-\mu)]
    	\int_{-\infty}^{\infty}dE\,\Phi_{xx}(E)A^2(E,\varepsilon)\\
		\label{eq:sigmaxy}\hspace{-0.7em}
		\sigma_{xy}&=B\hbar\int_{-\infty}^{\infty}d\varepsilon\,[-f'(\varepsilon-\mu)]
		\int_{-\infty}^{\infty}dE\,\Phi_{xy}(E)A^3(E,\varepsilon).
	\end{align}
The function $f'(\varepsilon-\mu)$ is the derivative of the Fermi-Dirac distribution, which is significant in an energy range of width $\sim\kB T$ around $\mu$ and reduces to $-\delta(\varepsilon-\mu)$ at zero temperature, thus solving the $\varepsilon$ integral in Eqs.~(\ref{eq:sigmaxx}) and (\ref{eq:sigmaxy}). If the transport functions are known either analytically or as fast piecewise functions, the evaluation of the conductivities at $T=0$ is reduced to a single quadrature depending parametrically on $\mu$ and $\Sigma(\mu)$. In practice, we bypass this quadrature by using, in the piecewise representation of $\Phi_{xx}(E)$ and $\Phi_{xy}(E)$, a set of mathematical functions for which the $E$-integral is known analytically. In this way, we obtain $\sigma_{xx}$ and $\sigma_{xy}$ accurately at $T=0$ without relying on any numerical quadrature. Likewise, the integral giving the DOS in Eq.~(\ref{eq:N}) is performed analytically in each piece of the piecewise function $N_0(E)$  \cite{Berthod_2025_Magnetotransport_piecewise}.

The Hall constant is finally deduced as
	\begin{equation}\label{eq:RH}
		\RH=\frac{\sigma_{xy}/B}{\sigma_{xx}^2},
	\end{equation}
which is known to approach $\RH^0=-1/(|e|n)$ when the scattering is sufficiently weak and the density sufficiently high \cite{Morpurgo_Giamarchi_2024_Hall_effect_low_density}. We solve these equations with a particular focus on the low-density regime for a microscopic model of self-energy describing electron scattering on weak pointlike impurities. Before introducing this model, we briefly discuss the limitations of the phenomenological self-energy model $\Sigma(\varepsilon)=-i\Gamma$.

\subsection{Drawbacks of the constant-scattering rate approximation}
\label{sec:Gamma}

\begin{figure}[tb]
\includegraphics[width=0.95\columnwidth]{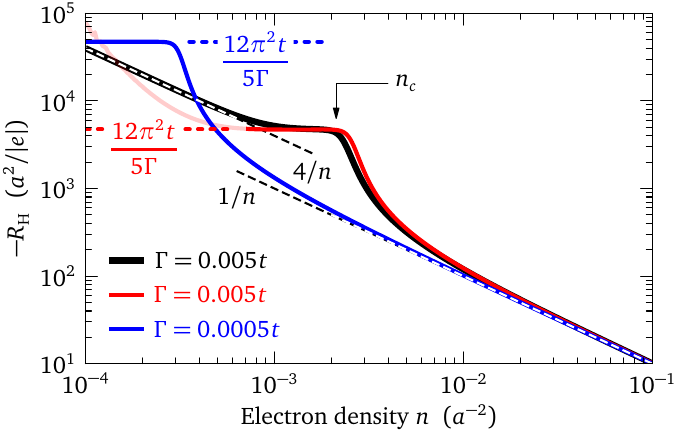}
\caption{\label{fig:Gamma}
Black line: zero-temperature Hall constant for a square lattice with scattering rate $\Gamma=0.005t$. The dashed lines indicate the asymptotic regimes $\RH=\RH^0$ and $\RH=4\RH^0$. $n_c$ shows Eq.~(20) of Ref.~\cite{Morpurgo_Giamarchi_2024_Hall_effect_low_density}. Red and blue lines: zero-temperature Hall constant for a continuum model with the same mass as the square lattice, a cutoff $E_c=80t$, and two values of the scattering rate. The dotted lines indicate the $\Gamma$-dependent plateau. 
}
\end{figure}

The Hall constant given by Eqs.~(\ref{eq:N0})--(\ref{eq:RH}) and its density and temperature dependences in the case of a constant scattering rate, $\Sigma(\varepsilon)=-i\Gamma$, were studied in Ref.~\cite{Morpurgo_Giamarchi_2024_Hall_effect_low_density} for a class of tight-binding models. One typical result for the dispersion $E_{\vec{k}}=-2t[\cos(k_xa)+\cos(k_ya)]$, corresponding to an isotropic square lattice with lattice parameter $a$ and hopping amplitude $t$, is reproduced as the black line in Fig.~\ref{fig:Gamma}. As the density decreases, $\RH$ crosses over at a density $n_c$, where the chemical potential traverses the lower band edge, from the semiclassical behavior $\RH^0$ to a regime where $\RH=4\RH^0$. While the semiclassical behavior is universal (independent of $E_{\vec{k}}$ and $\Gamma$), it turns out that the anomalous behavior $\RH=4\RH^0$ for $n\to0$ is not observed for all dispersion relations. 

To illustrate this, we show as red and blue lines in Fig.~\ref{fig:Gamma} the Hall constant computed in the continuum for a dispersion $E_{\vec{k}}=\hbar^2k^2/(2m)$. The mass was fixed to $m=\hbar^2/(2a^2t)$, such that both parabolic and cosine dispersions coincide at the band bottom. The right-hand side of Eq.~(\ref{eq:n}) is ultraviolet divergent for a constant scattering rate, which requires introducing a cutoff. We set the cutoff to $E_c=80t$, such that the bandwidth of the continuum model is ten times that of the square lattice. Both models display a departure from the semiclassical behavior as density decreases, followed by a $\Gamma$-dependent plateau. As seen in Eq.~(20) of Ref.~\cite{Morpurgo_Giamarchi_2024_Hall_effect_low_density}, the crossover density is proportional to $\Gamma$ and depends logarithmically on the bandwidth. In the parabolic case, $\RH$ stays at the plateau until the chemical potential reaches unphysically large negative values exceeding $-E_c$, where a crossover to $1/n^2$ behavior occurs (dimmed red color). The value $12\pi^2t/(5\Gamma)$ of $\RH$ on the plateau can be obtained by studying Eqs.~(\ref{eq:N0})--(\ref{eq:RH}) at $T=0$ in the regime $-E_c\ll\mu\ll0$ (see Sec.~\ref{sec:general_parabolic_dispersion}).

Despite the fact that the model $\Sigma(\varepsilon)=-i\Gamma$ gives a decent description at intermediate densities, one weakness is that the scattering rate fails to drop to zero at high and low energy, which has several peculiar consequences: (i) the spectral function is a Lorentzian with unbounded $1/\varepsilon^2$ tails; (ii) the interacting DOS extends to $\varepsilon=-\infty$ with $1/\varepsilon$ tail, reduced to $1/\varepsilon^2$ after introduction of the cutoff; (iii) the zero-temperature chemical potential diverges as $\mu\propto-1/n$ as $n\to 0$; and (iv) the conductivities depend on the ultraviolet structure of the theory, while in principle they should be set by low-energy properties at $T=0$. In the continuum model, the conductivities behave as $\sigma_{xx}\propto n^2/E_c^2$ and $\sigma_{xy}/B\propto n^4/(\Gamma E_c^4)$ on the plateau. This dependence on the cutoff cancels out in the Hall constant, but it nevertheless highlights a pathology of the phenomenological constant-scattering rate approximation. Self-energy functions originating from microscopic models should ensure that the scattering rate drops to zero at sufficiently high and low energies, which solves the problems (i)--(iv) mentioned above, but complicates the evaluation of the conductivities.

\subsection{Impurity scattering in the self-consistent Born approximation}
\label{sec:SCBA}

\subsubsection{Definition and general properties of the self-energy}

A concentration $n_i$ of identical impurities characterized by a potential $v(\vec{r})$ and located at random positions generates, after performing an average over the random positions, an effective self-energy for the electrons \cite{Book_Bruus_Flensberg_2016_Many_Body, Book_Berthod_2018_Spectroscopic_Probes}. The leading term (Born approximation) of this self-energy represents processes, where an electron scatters twice off the same impurity. It is given by 
	\begin{equation}\label{eq:BA}
		\Sigma^{\mathrm{BA}}(\vec{k},\varepsilon)=n_i\int\frac{d^2q}{(2\pi)^2}v(-\vec{q})
		G_0(\vec{k}+\vec{q},\varepsilon)v(\vec{q}),
	\end{equation}
with $v(\vec{q})$ the Fourier transform of the impurity potential and $G_0(\vec{k},\varepsilon)=1/(\varepsilon-E_{\vec{k}}+i0)$ the noninteracting Green's function describing the free propagation of the electron between these two scattering events. The self-energy changes the electron propagator from $G_0(\vec{k},\varepsilon)$ to $G(\vec{k},\varepsilon)=1/[\varepsilon-E_{\vec{k}}-\Sigma^{\mathrm{BA}}(\vec{k},\varepsilon)]$. The model can therefore be made self-consistent if one replaces $G_0$ by $G$ in Eq.~(\ref{eq:BA}).

In the following, we will use this self-consistent model in the case of pointlike impurities, for which $v(\vec{q})\equiv v_0$ is a constant. One sees that the self-energy loses any momentum dependence in this limit, and that the integral in Eq.~(\ref{eq:BA}) only samples the dispersion $E_{\vec{k}}$, which allows one to recast it as an energy integral involving the noninteracting DOS. We write this self-consistent model in the form
    \begin{equation}\label{eq:SCBA}
		\Sigma^{\mathrm{SCBA}}(\varepsilon)=(ga)^2\int_{-\infty}^{\infty}dE\,
		\frac{N_0(E)}{\varepsilon-E-\Sigma^{\mathrm{SCBA}}(\varepsilon)},
	\end{equation}
where $(ga)^2=n_iv_0^2/2$. The factor $1/2$ corrects for the spin sum, which is not present in Eq.~(\ref{eq:BA}) but included in $N_0(E)$. The parameter $g$ with units of energy will be our coupling constant, while $a$ is a model-dependent characteristic length that can be set as the unit of length when expressing $N_0(E)$. As it is derived from the Born approximation, the SCBA is valid in the limit of vanishing $v_0$, where the effect of impurity scattering only depends on the product $n_iv_0^2$. When we set the coupling $g$ to a finite value, it is understood that $n_i$ is large and $v_0$ is small, i.e., we are in a regime of dense and weak disorder. We show in Appendix~\ref{app:ED} that the SCBA is accurate in this regime, when checked against exact-diagonalization results. The accuracy of the local SCBA also implies that the vertex corrections of the conductivity are negligible, due to the decoupling of the two odd current vertices \cite{Khurana_1990_conductivity_infinite_d_Hubbard}. In the opposite regime of dilute and strong impurities, the self-energy depends separately on $n_i$ and $v_0$, becomes momentum dependent, and correspondingly the vertex corrections leading to Anderson localization become dominant.

The numerical evaluation of $\Sigma^{\mathrm{SCBA}}(\varepsilon)$ requires nesting a quadrature inside a root-finding algorithm. Here again, we bypass the quadrature by using, in the piecewise representation of $N_0(E)$, mathematical functions that allow us to perform the $E$-integral in Eq.~(\ref{eq:SCBA}) analytically in each piece. An illustration of the method for the cosine dispersion is given in Appendix~\ref{app:piecewise}. We then construct and store accurate piecewise representations of $\Sigma^{\mathrm{SCBA}}(\varepsilon)$ for each dispersion relation and each value of $g$. In this way, the inversion of Eq.~(\ref{eq:n}) for $\mu$, which is the most time-consuming part of the calculation, requires a single quadrature nested in a root-finding algorithm, while a straightforward implementation would have demanded two nested root findings involving up to seven nested quadratures.

For our purposes, the most important property that distinguishes qualitatively $\Sigma^{\mathrm{SCBA}}(\varepsilon)$ from the phenomenological model $\Sigma(\varepsilon)=-i\Gamma$ is that the scattering rate $-\mathrm{Im}\,\Sigma^{\mathrm{SCBA}}(\varepsilon)$ vanishes beyond an energy $\varepsilon_0$ outside the noninteracting band \footnote{As $-\mathrm{Im}\,\Sigma(\varepsilon)$ must remain strictly positive for causality reasons, an infinitesimal shift $i0$ must in principle be inserted into Eq.~(\ref{eq:SCBA}), such that $\mathrm{Im}\,\Sigma^{\mathrm{SCBA}}(\varepsilon)$ is never strictly zero.}. This may be seen by observing the behavior of Eq.~(\ref{eq:SCBA}) for $\varepsilon\to\infty$: If we assume that $\mathrm{Im}\,\Sigma^{\mathrm{SCBA}}(\varepsilon)$ on the right-hand side of the equation decays as $1/\varepsilon^{\nu}$ with $\nu\geqslant 0$ required by causality, we then see that $\mathrm{Im}\,\Sigma^{\mathrm{SCBA}}(\varepsilon)$ on the left-hand side would decay as $1/\varepsilon^{\nu+2}$. This excludes any solution with polynomial decay of $\mathrm{Im}\,\Sigma^{\mathrm{SCBA}}(\varepsilon)$ at infinity. A similar reasoning shows that $\mathrm{Re}\,\Sigma^{\mathrm{SCBA}}(\varepsilon)\sim1/\varepsilon$ at infinity.

For both the parabolic and cosine dispersions, the noninteracting DOS is discontinuous at the band bottom. In contrast, $\mathrm{Im}\,\Sigma^{\mathrm{SCBA}}(\varepsilon)$ approaches zero continuously like a square root at $\varepsilon_0$, as demonstrated in Appendix~\ref{app:edge}. On the other hand, $\mathrm{Re}\,\Sigma^{\mathrm{SCBA}}(\varepsilon)$ approaches a finite value at $\varepsilon_0$ and increases linearly for $\varepsilon>\varepsilon_0$. Therefore, using the notation $\Sigma^{\mathrm{SCBA}}(\varepsilon)=\Sigma_1(\varepsilon)+i\Sigma_2(\varepsilon)$, the asymptotic behavior of $\Sigma^{\mathrm{SCBA}}(\varepsilon)$ near the band bottom for $\varepsilon\gtrsim\varepsilon_0$ is
	\begin{equation}\label{eq:SCBA_asymptotic}
		\Sigma^{\mathrm{SCBA}}(\varepsilon)=\Sigma_1(\varepsilon_0)+
		\Sigma_1'(\varepsilon_0)(\varepsilon-\varepsilon_0)-iC\sqrt{\varepsilon-\varepsilon_0}.
	\end{equation}
An expression for the constant $C$ is provided in Appendix~\ref{app:edge}.

The SCBA offers a peculiar property that proves useful in obtaining analytical results for the zero-temperature conductivities: The interacting DOS is proportional to the scattering rate. This may be seen from Eqs.~(\ref{eq:SCBA}), (\ref{eq:A}), and (\ref{eq:N}), using the fact that $\mathrm{Im}\,\{1/[\varepsilon-E-\Sigma(\varepsilon)]\}=-\pi\,A(E,\varepsilon)$:
	\begin{equation}\label{eq:SCBA_DOS}
		N(\varepsilon)=\frac{-\mathrm{Im}\,\Sigma^{\mathrm{SCBA}}(\varepsilon)}{\pi (ga)^2}.
	\end{equation}
If the asymptotic form $N(\varepsilon\gtrsim\varepsilon_0)=\frac{C}{\pi(ga)^2}\sqrt{\varepsilon-\varepsilon_0}$ is injected in Eq.~(\ref{eq:n}) and the temperature is set to zero, the density is found to vary as $n=\frac{2C}{3\pi(ga)^2}(\mu-\varepsilon_0)^{3/2}$, which provides the asymptotic low-density behavior of the chemical potential at $T=0$:
	\begin{equation}\label{eq:mu}
		\mu=\varepsilon_0+\left(\frac{3\pi(ga)^2}{2C}n\right)^{2/3}
		\qquad(T=0, n\to0).
	\end{equation}
This is qualitatively different from the case of a constant scattering rate, where the chemical potential diverges as $-\Gamma/n$ at low carrier density \cite{Morpurgo_Giamarchi_2024_Hall_effect_low_density}. Equation~(\ref{eq:mu}) shows that the energy $\varepsilon_0$ and the amplitude $C$ are the key properties of the self-energy that control the asymptotic low-density physics. In the next subsections, we display the dependence of these two parameters on the coupling constant $g$ for the parabolic and cosine dispersions.

\subsubsection{SCBA self-energy for a parabolic dispersion}
\label{sec:k2-SCBA}

When using the parabolic model $E_{\vec{k}}=\hbar^2k^2/(2m)$, we set the Bohr radius corresponding to the mass $m$ as the unit of length, i.e., $a=4\pi\epsilon_0\hbar^2/(me^2)$. For the unit of energy, we use the corresponding Hartree $\mathrm{Hr}=\hbar^2/(ma^2)$, which is twice the kinetic energy for a wavevector $1/a$. The DOS as defined by Eq.~(\ref{eq:N0}) is flat for positive energies, $N_0(E)=N_0\theta(E)$, where $N_0=m/(\pi\hbar^2)=1/(\pi a^2\mathrm{Hr})$ and $\theta(E)$ is the Heaviside step function. Equation~(\ref{eq:SCBA}) is logarithmically divergent for this unbounded $N_0(E)$. We therefore introduce an ultraviolet cutoff $E_c$ for the noninteracting DOS of the parabolic band,
	\begin{equation}\label{eq:N0_k2}
		N_0(E)=N_0\theta(E)\theta(E_c-E).
	\end{equation}
It follows that the SCBA self-energy is given by the implicit equation
	\begin{multline}\label{eq:SCBA_parabolic}
		\Sigma^{\mathrm{SCBA}}(\varepsilon)=(ga)^2N_0\left\{
		\ln\left[\varepsilon-\Sigma^{\mathrm{SCBA}}(\varepsilon)\right]\right.\\\left.
		-\ln\left[\varepsilon-\Sigma^{\mathrm{SCBA}}(\varepsilon)-E_c\right]\right\}.
	\end{multline}
Despite the cutoff dependence of the self-energy, we will see that the conductivities and the Hall coefficient have a well-defined limit when the cutoff is sent to infinity.

\begin{figure}[tb]
\includegraphics[width=0.95\columnwidth]{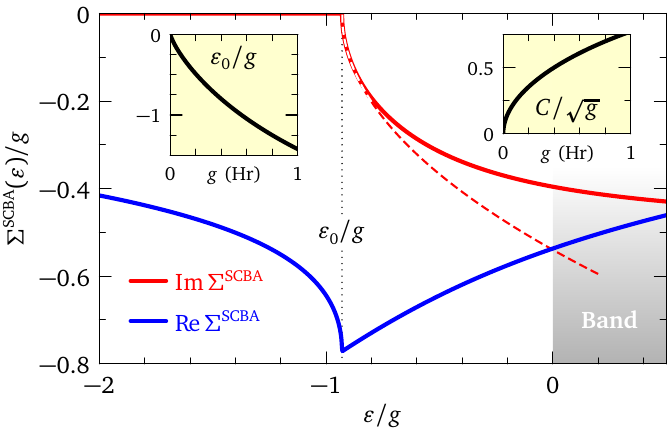}
\caption{\label{fig:SCBA_k2}
Real part (blue) and imaginary part (red) of the self-energy given by Eq.~(\ref{eq:SCBA_parabolic}) for $g=0.5$~Hr and $E_c=10$~Hr. The dashed line shows the imaginary part of Eq.~(\ref{eq:SCBA_asymptotic}) with $C$ given by Eq.~(\ref{eq:C_parabolic}). The shaded region indicates the noninteracting band. Insets: illustration of the dependences of $\varepsilon_0$ and $C$ on $g$, as given by Eq.~(\ref{eq:parabolic}).
}
\end{figure}

The numerical solution of Eq.~(\ref{eq:SCBA_parabolic}) for $g=0.5$~Hr and $E_c=10$~Hr is depicted in Fig.~\ref{fig:SCBA_k2}. A large coupling $g$ is chosen here for illustration purposes, while small values will be considered later to generate realistic conductivities. As expected, the imaginary part vanishes as a square root at a coupling-dependent energy $\varepsilon_0$ below the edge of the noninteracting band. The real part is continuous at $\varepsilon_0$ with a vertical slope below $\varepsilon_0$ and a finite slope above $\varepsilon_0$ (Appendix~\ref{app:edge}). The relative simplicity of Eq.~(\ref{eq:SCBA_parabolic}) allows us to obtain closed expressions for $\varepsilon_0$ and $C$ in this case (Appendix~\ref{app:edge}), which we express here using the dimensionless parameter $\gamma=4(ga)^2N_0/E_c$:
	\begin{subequations}\label{eq:parabolic}\begin{align}
		\label{eq:eps0_parabolic}
		\varepsilon_0&=-\frac{E_c}{4}\left[2\left(\sqrt{1+\gamma}-1\right)
		+\gamma\ln\frac{\sqrt{1+\gamma}+1}{\sqrt{1+\gamma}-1}\right]\\
		\label{eq:C_parabolic}
		C&=\frac{\sqrt{E_c\gamma/2}}{(1+\gamma)^{1/4}}.
	\end{align}\end{subequations}
These formula are displayed in the insets of Fig.~\ref{fig:SCBA_k2}. $\varepsilon_0$ varies initially as $g^2$ with a logarithmic correction, while $C$ grows initially as $g$. Note that $C$ approaches $(2/\pi)^{1/2}g/\sqrt{\mathrm{Hr}}$ for a large cutoff, while $\varepsilon_0$ depends logarithmically on the cutoff as $-g^2/(\pi \mathrm{Hr})\{1-\ln[g^2/(\pi \mathrm{Hr}E_c)]\}$.

\subsubsection{SCBA self-energy for a cosine dispersion}
\label{sec:cosk-SCBA}

In the tight-binding approximation, a two-dimensional square lattice with nearest-neighbor hopping energy $-t$ has a dispersion relation $E_{\vec{k}}=-2t[\cos(k_xa)+\cos(k_ya)]$, where $a$ is the lattice parameter. We take $t>0$ and insert a minus sign, such that the bottom of the band is at $\vec{k}=\vec{0}$ with a positive curvature and a mass given by $m=\hbar^2/(2a^2t)$. When using this model, we choose $a$ as the unit of length and $t$ as the unit of energy. The exact noninteracting DOS is known in terms of an elliptic function. As this elliptic function prevents the analytical evaluation of the $E$-integrals in Eqs.~(\ref{eq:N}) and (\ref{eq:SCBA}), we use a piecewise representation of $N_0(E)$ to avoid numerical quadratures (see Appendix~\ref{app:piecewise}).

\begin{figure}[tb]
\includegraphics[width=0.95\columnwidth]{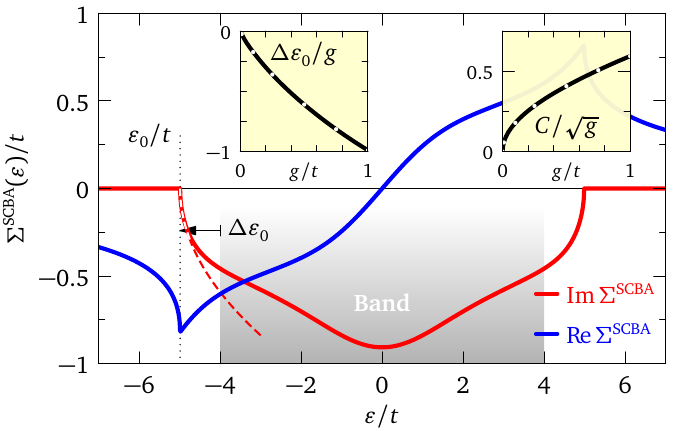}
\caption{\label{fig:SCBA_cosk}
Real part (blue) and imaginary part (red) of the self-energy given by Eq.~(\ref{eq:SCBA}) for a cosine dispersion and a coupling $g=t$. The dashed line shows the imaginary part of Eq.~(\ref{eq:SCBA_asymptotic}) with $C$ given by Eq.~(\ref{eq:C}). The shaded region indicates the noninteracting band. Insets: illustration of the dependences of $\varepsilon_0$ and $C$ on $g$. $\Delta\varepsilon_0=\varepsilon_0+4t$ is the energy shift from the band edge.
}
\end{figure}

The numerical solution of Eq.~(\ref{eq:SCBA}) for $g=t$ is depicted in Fig.~\ref{fig:SCBA_cosk}. Again, we use a large coupling for illustration purposes. We show the full energy dependence to illustrate the particle-hole symmetry resulting from this dispersion, with odd and even real and imaginary parts, respectively. The square-root behavior of the imaginary part, the vertical slope of the real part for $\varepsilon<\varepsilon_0$, and the finite slope for $\varepsilon>\varepsilon_0$ are similar to the results found in the parabolic case. The noninteracting DOS has edge discontinuities at $\varepsilon=\pm4t$ and a van Hove singularity at $\varepsilon=0$ (Appendix~\ref{app:piecewise}). These nonanalyticities are renormalized by impurity scattering, leading to shifted and rounded features in $N(\varepsilon)\propto-\mathrm{Im}\,\Sigma^{\mathrm{SCBA}}(\varepsilon)$. The figure also shows the dependences of $\varepsilon_0$ and $C$ on the coupling $g$, where it is seen that the behaviors are similar to those observed with the parabolic band.

\section{Numerical and analytical asymptotic results at zero temperature}
\label{sec:results}

A glance at Eqs.~(\ref{eq:sigmaxx}), (\ref{eq:sigmaxy}), and (\ref{eq:A}) reveals that, at $T=0$, the conductivities only depend on the chemical potential and self-energy through the complex number $\mu-\Sigma(\mu)$. For a parabolic band, the transport functions $\Phi_{xx}(E)$ and $\Phi_{xy}(E)$ are known analytically and explicit functions of $\mu-\Sigma(\mu)$ can be deduced for $\sigma_{xx}$, $\sigma_{xy}$, and $\RH$. We start by providing these functions, as they are valid for any local self-energy and may be used, in particular, to calculate the values of $\RH$ on the plateau, as displayed in Fig.~\ref{fig:Gamma}. We then present numerical and analytical asymptotic results obtained with the SCBA self-energy, both for the quadratic and cosine dispersions. Like in the constant-scattering rate approximation, we find a low-density quantum regime, where the conductivities and Hall constant deviate from the semiclassical behavior. Unlike in this approximation, however, the deviations present identical universal features for both dispersion relations.

\subsection{Zero-temperature conductivity tensor for a parabolic dispersion and a generic local self-energy}
\label{sec:general_parabolic_dispersion}

At zero temperature, Eqs.~(\ref{eq:sigmaxx}) and (\ref{eq:sigmaxy}) reduce to
	\begin{align}
		\label{eq:sigmaxx_T=0}
		\sigma_{xx}&=\pi\hbar\int_{-\infty}^{\infty}dE\,\Phi_{xx}(E)\left[
		\mathrm{Im}\,\frac{-1/\pi}{\mu-\Sigma(\mu)-E}\right]^2\\
		\label{eq:sigmaxy_T=0}
		\sigma_{xy}&=B\hbar\int_{-\infty}^{\infty}dE\,\Phi_{xy}(E)\left[
		\mathrm{Im}\,\frac{-1/\pi}{\mu-\Sigma(\mu)-E}\right]^3.
	\end{align}
For a parabolic dispersion, both transport functions are linear functions of the energy:
	\begin{align}
		\Phi_{xx}(E)&=\left(\frac{e}{\hbar}\right)^2\frac{E}{\pi}\theta(E)\\
		\Phi_{xy}(E)&=-\left(\frac{|e|}{\hbar}\right)^3\frac{2\pi\hbar^2E}{3m}\theta(E).
	\end{align}
The $E$-integrals in Eqs.~(\ref{eq:sigmaxx_T=0}) and (\ref{eq:sigmaxy_T=0}) are therefore convergent without ultraviolet cutoff. It is convenient to introduce the modulus and the phase of the complex number $\mu-\Sigma(\mu)\equiv Me^{i\phi}$. Since $\mathrm{Im}\,\Sigma(\mu)$ is strictly negative, the phase is constrained to the interval $0<\phi<\pi$. Evaluating the integrals, we find the conductivities in terms of those variables,
	\begin{align}
		\label{eq:sigmaxx_M_phi}
		\sigma_{xx}&=\frac{e^2}{h}\frac{1+(\pi-\phi)\cot\phi}{\pi}\\
		\label{eq:sigmaxy_M_phi}
		\sigma_{xy}&=-\frac{e^2}{h}\frac{Ba^2}{\Phi_0}\frac{\mathrm{Hr}}{M}
		\frac{1+(\pi-\phi)\cot\phi-\frac{1}{3}\sin^2\phi}{\sin\phi},
	\end{align}
where $\Phi_0=h/|e|$ is the magnetic flux quantum and we recall that $\mathrm{Hr}=\hbar^2/(ma^2)$.

Interestingly, $\sigma_{xx}$ only depends on the phase $\phi$. Equation~(\ref{eq:sigmaxx_M_phi}) recovers the Drude formula in the semiclassical limit, where the chemical potential is given by the electron gas formula, $\mu=\pi\hbar^2n/m$, and the self-energy is related to the carrier lifetime $\tau$ as $\Sigma(\mu)=-i\hbar/(2\tau)$. For large $\tau$, the phase approaches zero as $\phi\to\hbar/(2\tau\mu)$ and Eq.~(\ref{eq:sigmaxx_M_phi}) gives $\sigma_{xx}=e^2/(h\phi)=ne^2\tau/m$. Likewise, $\sigma_{xy}$ approaches the Drude result $-ne^2\omega_c\tau^2/m$ in this limit, where $\omega_c=|e|B/m$ is the cyclotron frequency.

Inserting Eqs.~(\ref{eq:sigmaxx_M_phi}) and (\ref{eq:sigmaxy_M_phi}) in Eq.~(\ref{eq:RH}), we obtain the Hall constant
	\begin{subequations}\label{eq:RH_M_phi}\begin{align}
		\RH&=-\frac{1}{|e|N_0}\frac{F(\phi)}{M}\\
		F(\phi)&=\pi\frac{(\pi-\phi)\cos\phi+\frac{3}{4}\sin\phi+\frac{1}{12}\sin(3\phi)}
		{[(\pi-\phi)\cos\phi+\sin\phi]^2}.
	\end{align}\end{subequations}
Since $F(0)=1$, this expression recovers the semiclassical result $\RH^0$ if $\mu-\Sigma(\mu)=n/N_0+i0$. Another important limit is $\phi\to\pi$, which occurs in the quantum regime at vanishingly low density, when the renormalized chemical potential $\mu-\mathrm{Re}\,\Sigma(\mu)$ is below the noninteracting band and its magnitude is much larger than the scattering rate $-\mathrm{Im}\,\Sigma(\mu)$. In this limit, $F(\phi)$ diverges as $\frac{6\pi}{5}\frac{1}{\pi-\phi}$. For instance, if $\Sigma(\varepsilon)=-i\Gamma$ and $n\to0$, $\mu$ approaches $-\infty$, $\phi$ approches $\pi+\Gamma/\mu$, and $M$ approaches $-\mu$, such that $\RH$ approaches $-6\pi^2/(5\Gamma)$ in units of $a^2\mathrm{Hr}/|e|$. This becomes $12\pi^2t/(5\Gamma)$ with the units chosen in Fig.~\ref{fig:Gamma}.

\subsection{Anomalous power laws at low density in the self-consistent Born approximation}

\subsubsection{Parabolic dispersion}

\begin{figure}[tb]
\includegraphics[width=\columnwidth]{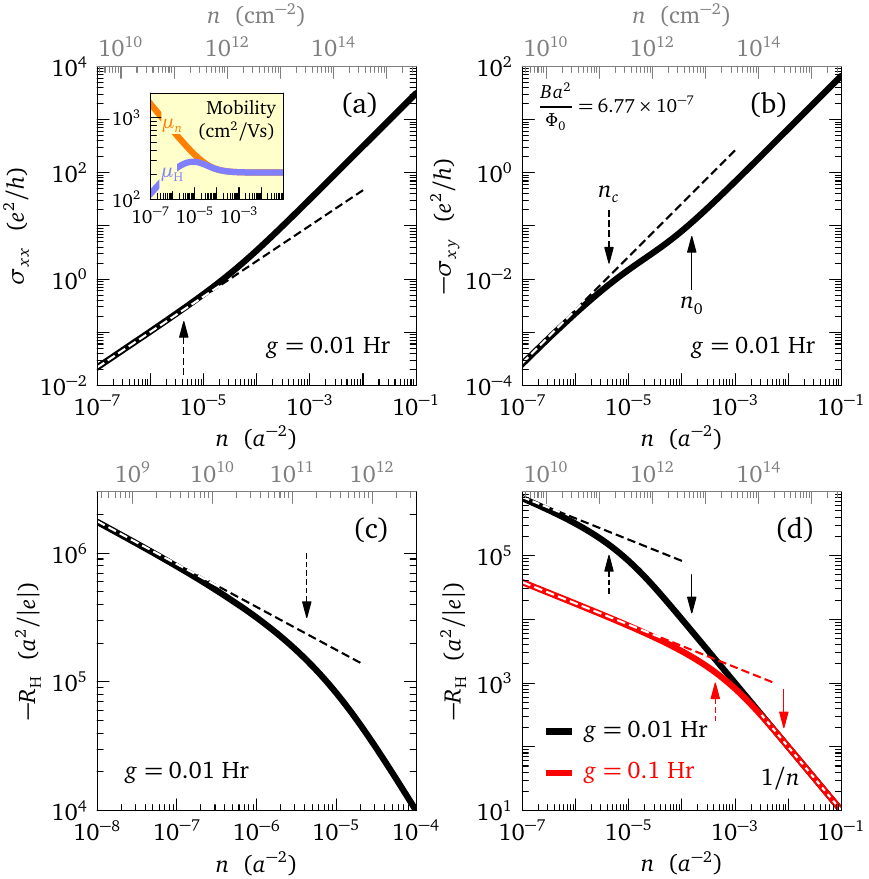}
\caption{\label{fig:k2-SCBA}
(a) Normal conductivity, (b) Hall conductivity, and (c) Hall constant vs density at $T=0$ for a parabolic band and the SCBA self-energy with $g=0.01$~Hr. (d) Comparison of the Hall constants for two values of $g$. All quantities are computed with a cutoff $E_c=10$~Hr and $\sigma_{xy}$ is displayed for a field $B=1$~T and a mass $m$ equal to the electron mass. The dashed lines show the asymptotic results given in the text. The solid and dashed arrows show the density $n_0$ at which $\mu=0$ in Eq.~(\ref{eq:mu}) and $n_c$ from Eq.~(\ref{eq:nc-k2}), respectively. Inset: mobilities $\mu_n=\sigma_{xx}/(n|e|)$ and $\mu_{\mathrm{H}}=-\sigma_{xx}\RH$ for the same density range as in panel (a).}
\end{figure}

In Figs.~\ref{fig:k2-SCBA}(a) and \ref{fig:k2-SCBA}(b), we plot the normal and Hall conductivities computed with Eqs.~(\ref{eq:N0})--(\ref{eq:RH}) at $T=0$ for a range of densities, using a parabolic dispersion and the SCBA self-energy (Sec.~\ref{sec:k2-SCBA}) with $g=0.01$~Hr. Since the self-energy in Eq.~(\ref{eq:SCBA_parabolic}) is not convergent without a cutoff imposed to $N_0(E)$, we apply the same cutoff procedure to $\Phi_{xx}(E)$ and $\Phi_{xy}(E)$ for consistency in the numerical calculations, although this has no visible effect on the conductivities displayed in the figure. To express the densities in cm$^{-2}$ as indicated on the upper axis of the graphs, we set the band mass to the bare electron mass, such that $a$ coincides with the usual Bohr radius, Hr with the usual Hartree, and $g\approx0.27$~eV.

With reducing the density $n$, both conductivities quit the semiclassical behavior proportional to $n$. This occurs at a density close to the value where the chemical potential crosses the bottom of the noninteracting band, as estimated by $n_0$, the solution of Eq.~(\ref{eq:mu}) with $\mu=0$. For the particular value of $g$, the change of behavior takes place at physically accessible densities of order $10^{12}$~cm$^{-2}$, when the conductivity is similar to the quantum $e^2/h$ and the mobility is of order 200~cm$^2$/Vs. We compare in the inset of Fig.~\ref{fig:k2-SCBA} two different definitions of the mobility, $\mu_n=\sigma_{xx}/(n|e|)$ and $\mu_{\mathrm{H}}=-\sigma_{xx}\RH$, which coincide in the semiclassical regime. In the low-density quantum regime, $\sigma_{xx}$ displays a power law with exponent $2/3$. Accordingly, the mobilities $\mu_n$ and $\mu_{\mathrm{H}}$ cease to be good measures of the scattering. On the other hand, $\sigma_{xy}$ crosses over from the high-density linear regime to a low-density regime where it is still linear in $n$, but with a different factor. Consequently, the Hall constant crosses over from the semiclassical $\sim 1/n$ to an anomalous behavior $\sim 1/n^{1/3}$, as illustrated in Figs.~\ref{fig:k2-SCBA}(c) and \ref{fig:k2-SCBA}(d). The crossover density $n_c$ [defined in Eq.~(\ref{eq:nc-k2})], below which $\RH$ is no longer a good measure of the number of carriers, increases like $g^2$ as the scattering gets stronger.

The low-density power laws plotted as dashed lines in Fig.~\ref{fig:k2-SCBA} may be derived from Eq.~(\ref{eq:sigmaxx_M_phi})--(\ref{eq:RH_M_phi}). Using Eqs.~(\ref{eq:Sigma1_eps0}), (\ref{eq:Sigma2_eps0}), and (\ref{eq:mu}), we find that the asymptotic expression for $\mu-\Sigma^{\mathrm{SCBA}}(\mu)$ when $n$ approches zero is
	\begin{equation}\label{eq:M_phi}
		Me^{i\phi}=-\frac{E_c}{2}\left(\sqrt{1+\gamma}-1\right)
		+iC\left(\frac{3\pi(ga)^2}{2C}n\right)^{1/3},
	\end{equation}
where the cutoff dependence arises from the self-energy. This expression shows that $M\cos\phi$ is negative as $n\to0$, such that $\phi$ approaches $\pi$ in this limit. Inserting Eq.~(\ref{eq:M_phi}) into Eq.~(\ref{eq:sigmaxx_M_phi}) and letting the cutoff $E_c$ go to infinity in the resulting expression, we find
	\begin{equation}\label{eq:sigmaxx-k2}
		\sigma_{xx}=\frac{e^2}{h}\frac{\pi}{3^{1/3}}
		\left(\frac{g}{\mathrm{Hr}}\right)^{-\frac{4}{3}}\left(na^2\right)^{\frac{2}{3}}
		\qquad(n\to0).
	\end{equation}
This formula shows that if the quantities $g^2$ and $n$ have similar values in atomic Hartree units, then the conductivity is of the order of $e^2/h$. For the Hall conductivity, we find the linear-in-$n$ behavior
	\begin{equation}\label{eq:sigmaxy-k2}
		\sigma_{xy}=-\frac{e^2}{h}\frac{Ba^2}{\Phi_0}\frac{2\pi^4}{5}
		\left(\frac{g}{\mathrm{Hr}}\right)^{-4}na^2
		\qquad(n\to0).
	\end{equation}
Since $\sigma_{xy}\sim g^{-4}$ and $\sigma_{xx}\sim g^{-4/3}$, the information on the coupling constant $g$ does not disappear from $\RH$. Indeed, for the Hall constant we find
	\begin{equation}
		\label{eq:RH-k2}
		\RH=-\frac{a^2}{|e|}\frac{3^{2/3}2\pi^2}{5}
		\left(\frac{g}{\mathrm{Hr}}\right)^{-\frac{4}{3}}(na^2)^{-\frac{1}{3}}
		\qquad(n\to0).
	\end{equation}
While $n_0$ marks the onset of the deviations from the semiclassical regime, a characteristic density $n_c$ marking the onset of the quantum regime may be defined as the crossing point between Eq.~(\ref{eq:RH-k2}) and the semiclassical expression $\RH^0$. We thus get
	\begin{equation}\label{eq:nc-k2}
		n_c=\frac{1}{a^2}\frac{5^{3/2}}{6\sqrt{2}\pi^3}\left(\frac{g}{\mathrm{Hr}}\right)^2,
	\end{equation}
which is displayed as dashed arrows in Fig.~\ref{fig:k2-SCBA}.

\subsubsection{Cosine dispersion}

\begin{figure}[tb]
\includegraphics[width=\columnwidth]{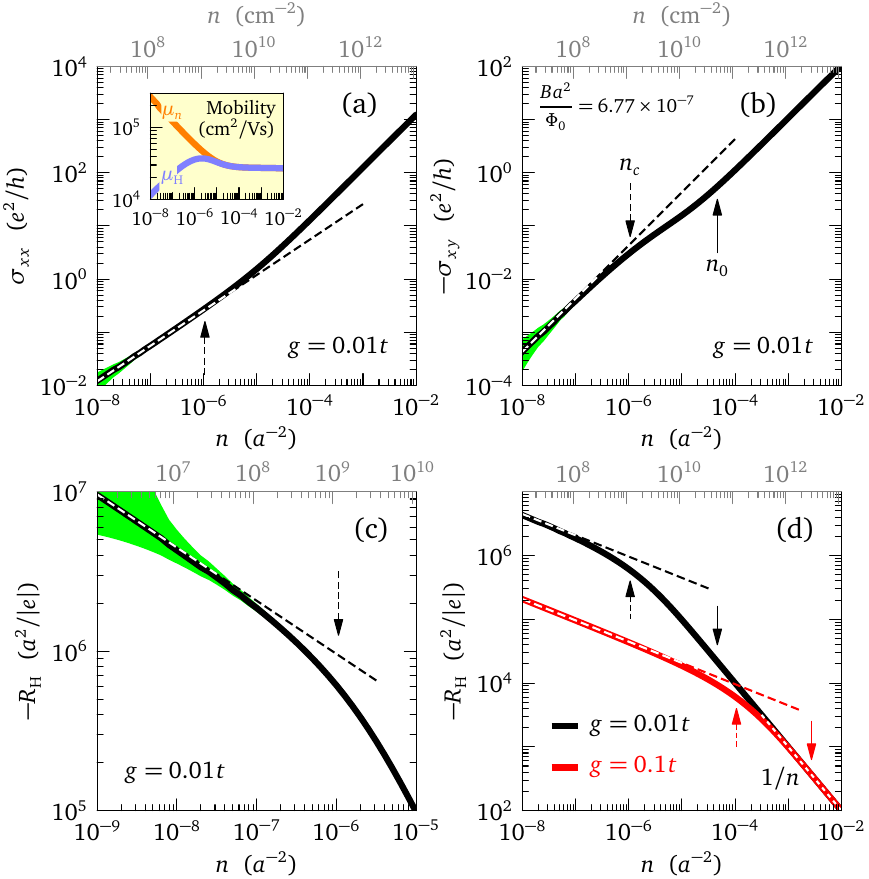}
\caption{\label{fig:cosk-SCBA}
Same as Fig.~\ref{fig:k2-SCBA} for a cosine dispersion. $\sigma_{xy}$ is displayed for $B=0.031$~T and $a=3$~\AA{}. The dashed arrows show $n_c$ from Eq.~(\ref{eq:nc-cosk}). The green shade shows the effect of changing $\mu$ to $\mu\times(1\pm10^{-7})$ in the numerical calculations.
}
\end{figure}

As the conductivities are controlled by $\mu-\Sigma(\mu)$ [Eqs.~(\ref{eq:sigmaxx_T=0}) and (\ref{eq:sigmaxy_T=0})] and both $\Sigma(\varepsilon)$ and $N(\varepsilon)$ display universality at low density in the SCBA [Eqs.~(\ref{eq:SCBA_asymptotic}) and (\ref{eq:SCBA_DOS})], we expect universality in the conductivities, at least  for all dispersion relations that yield linear-in-energy transport functions at low energy. This is the case if the bottom of the band is parabolic, like in the cosine dispersion. The conductivities and Hall constant obtained at $T=0$ using a cosine dispersion and the corresponding SCBA self-energy (Sec.~\ref{sec:cosk-SCBA}) with $g=0.01t$ are plotted in Fig.~\ref{fig:cosk-SCBA}. A comparison with $g=0.1t$ is shown in Fig.~\ref{fig:cosk-SCBA}(d). The lattice parameter is set to $a=3$~\AA{} in order to express the densities in cm$^{-2}$ and evaluate the mobilities, and we plot the Hall conductivity for a field $B=0.031$~T in order to have the same magnetic flux as with the parabolic band. We find striking similarities with the parabolic case and, in particular, the same anomalous power-law exponents for $\sigma_{xx}$ and $\RH$ in the quantum regime.

One can see that the numerical data present some noise below $na^2\sim10^{-7}$. It turns out that the calculations are challenging in this regime, as very small variations in the value of $\mu$ have significant consequences for $\sigma_{xx}$, $\sigma_{xy}$, and $\RH$. The green-shaded regions in Fig.~\ref{fig:cosk-SCBA} represent the variations in the value of $\sigma_{xx}$, $\sigma_{xy}$, and $\RH$ if a \emph{relative} change of only $\pm10^{-7}$ is applied to $\mu$. This illustrates the interest of having analytical results for benchmarking the numerical calculations.

To obtain asymptotic expressions for the conductivities, we linearize the transport functions in Eqs.~(\ref{eq:sigmaxx_T=0}) and (\ref{eq:sigmaxy_T=0}), which leads to expressions analogous to Eqs.~(\ref{eq:sigmaxx_M_phi}) and (\ref{eq:sigmaxy_M_phi}). The linearized transport functions are \cite{Morpurgo_Giamarchi_2024_Hall_effect_low_density} $\Phi_{xx}(E)=\left(e/\hbar\right)^2(1/\pi)(E+4t)\theta(E+4t)$ and $\Phi_{xy}(E)=-\left(|e|/\hbar\right)^3a^2t(4\pi/3)(E+4t)\theta(E+4t)$. We then use Eqs.~(\ref{eq:SCBA_asymptotic}) and (\ref{eq:mu}) to determine $\mu-\Sigma^{\mathrm{SCBA}}(\mu)$, and we expand for $\mu$ close to $\varepsilon_0$. The result is
	\begin{align}
		\sigma_{xx}&=\frac{e^2}{h}\frac{1}{(12\pi)^{1/3}}\frac{\left(Cg\right)^{4/3}}
		{\left[\varepsilon_0-\Sigma_1(\varepsilon_0)+4t\right]^2}\left(na^2\right)^{\frac{2}{3}}\\
		\sigma_{xy}&=-\frac{e^2}{h}\frac{Ba^2}{\Phi_0}\frac{2\pi}{5}\frac{t(Cg)^2}
		{\left[\varepsilon_0-\Sigma_1(\varepsilon_0)+4t\right]^4}na^2.
	\end{align}
These expressions depend on $\varepsilon_0$ and $C$, for which we lack closed forms in the cosine case. The dependence on $\varepsilon_0$ drops from $\RH$, which only depends on $C$:
	\begin{equation}\label{eq:RH-cosk}
		\RH=-\frac{a^2}{|e|}\frac{4(18\pi^5)^{1/3}}{5}\frac{t}{(Cg)^{2/3}}(na^2)^{-\frac{1}{3}}.
	\end{equation}
We can again define the crossover density $n_c$ displayed in Fig.~\ref{fig:cosk-SCBA} by comparing Eq.~(\ref{eq:RH-cosk}) with $\RH^0$:
	\begin{equation}\label{eq:nc-cosk}
		n_c=\frac{1}{a^2}\frac{5^{3/2}}{24\sqrt{2}\pi^{5/2}}\frac{Cg}{t^{3/2}}.
	\end{equation}
Since $C\propto g$ (Fig.~\ref{fig:SCBA_cosk}), $n_c$ varies as $g^2$, like in the parabolic case.

\section{Discussion}
\label{sec:discussion}

The results presented in the preceding section show that marked deviations from the semiclassical transport behavior do occur in a model of disordered metal at low carrier density. In this discussion, we first describe the possible manifestations of these deviations in field-effect experiments performed on two-dimensional materials and we propose strategies to reveal and analyze them. We then comment on probable limitations in the range of validity of the model, due to localization of the carriers by the disorder.

\subsection{Signatures of the quantum regime in field-effect experiments}

\begin{figure}[tb]
\includegraphics[width=\columnwidth]{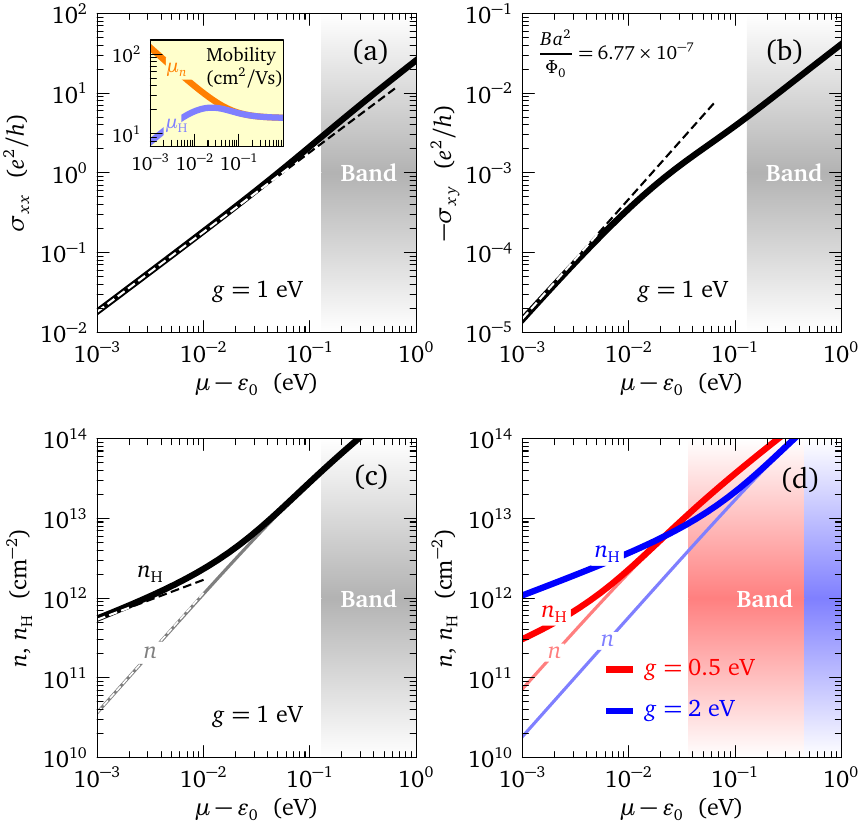}
\caption{\label{fig:Hall_SCBA_mu}
(a) Normal conductivity, (b) Hall conductivity, and (c) Hall and true carrier densities vs chemical potential at $T=0$ for a parabolic band and the SCBA self-energy with $g=1$~eV. $\mu>0$ in the shaded regions. (d) Comparison of the Hall and true carrier densities for two values of $g$. All quantities are computed with a cutoff $E_c=10$~Hr and a mass $m$ equal to the electron mass. $\sigma_{xy}$ is displayed for a field $B=1$~T. The dashed lines show the asymptotic results given in the text. Inset: mobilities $\mu_n=\sigma_{xx}/(n|e|)$ and $\mu_{\mathrm{H}}=-\sigma_{xx}\RH$ for the same range of chemical potential as in panel (a).
}
\end{figure}

Two-dimensional metals with low carrier densities were realized experimentally by field-effect doping of 2D semiconductors \cite{Yamagishi_Takeya_2010_Hall_organic_thin_film, Uemura_Takeya_2012_Temperature_dependance_Hall, Pradhan_Balicas_2015_weak_field_Hall_TMD, Masseroni_Duprez_2024_TMD_graphene_Spin_Orbit}. The external electric field pushes the chemical potential inside the conduction or valence band of the semiconductor, providing electron or hole carriers. In those experiments, the number of doped carriers is commonly estimated using the Hall number
    \begin{equation}
		\nH=-\frac{1}{|e|\RH}.
	\end{equation}
Given that $\RH$ departs from the semiclassical behavior towards lower values at low density (Figs.~\ref{fig:k2-SCBA} and \ref{fig:cosk-SCBA}), we expect that the Hall number overestimates the carrier density in this regime. The control variable of field-effect experiments is the gate voltage, which should be related linearly with the chemical potential in the absence of nonlinearities of the dielectric response. In Fig.~\ref{fig:Hall_SCBA_mu}, we therefore plot the conductivities and the Hall constant for the parabolic dispersion, using the chemical potential rather than the density as the abscissa. We set $m$ to the electron mass to express the chemical potential in eV and we use a coupling constant $g=1$~eV. With these parameters, the normal conductivity reaches $\sim e^2/h$ when the chemical potential is 50~meV above the interacting band edge $\varepsilon_0$, and the mobility is $\sim 20$~cm$^2$/Vs [Fig.~\ref{fig:Hall_SCBA_mu}(a)]. These values are similar to those typically observed in experiments \cite{Yamagishi_Takeya_2010_Hall_organic_thin_film, Uemura_Takeya_2012_Temperature_dependance_Hall, Pradhan_Balicas_2015_weak_field_Hall_TMD}.

Interestingly, $\sigma_{xx}$ is seemingly linear as a function of $\mu$, in contrast to what is observed versus $n$ in Fig.~\ref{fig:k2-SCBA}(a). A linear variation of $\sigma_{xx}$ is commonly observed in field-effect experiments, where it is used to determine the band edge by extrapolation \cite{Review_Gutierrez_Morpurgo_2021_technique_measure_band_gaps}. In the quantum regime, the linearity arises because $\sigma_{xx}\propto n^{2/3}$, while $n\propto(\mu-\varepsilon_0)^{3/2}$. The Hall conductivity $\sigma_{xy}$, which is proportional to $n$ [Fig.~\ref{fig:k2-SCBA}(b)], changes from $(\mu-\varepsilon_0)^{3/2}$ in the quantum regime to $(\mu-\varepsilon_0)^1$ in the semiclassical regime, which, if experimentally observed, would provide evidence for the quantum regime. The Hall mobility provides yet another signature of the crossover to the quantum regime, in the form of a local maximum. We note that a maximum in the gate-voltage dependence of the Hall mobility of WSe$_2$ field-effect transistors has been reported in Ref.~\cite{Pradhan_Balicas_2015_weak_field_Hall_TMD}. In Figs.~\ref{fig:Hall_SCBA_mu}(c) and \ref{fig:Hall_SCBA_mu}(d), we compare $\nH$ with the actual carrier density $n$. When the chemical potential is between $\varepsilon_0$ (bottom of the ``impurity band'') and $0$ (bottom of the noninteracting band), the Hall carrier density may indeed overestimate the true carrier density by several orders of magnitude. In Ref.~\cite{Yamagishi_Takeya_2010_Hall_organic_thin_film}, clear deviations of $1/\RH$ from the semiclassical linear behavior were reported for two organic field-effect transistors.

A complementary approach to reveal the transition into the quantum regime is to study scaling relations displayed by conductivities. In the semiclassical regime, two remarkable universalities have proven very important in analyzing experimental transport data: $\RH$ is independent of the scattering and the mobility $\mu_{\mathrm{H}}=-\sigma_{xx}\RH$ is independent of the carrier density. These two properties hold true, irrespective of the microscopic origin of the scattering. In the quantum regime, it is generally possible that suitable combinations of $\sigma_{xx}$ and $\RH$ can be constructed, one depending only on the density and the other depending only on the coupling constant. It is unlikely that these combinations are identical for all scattering mechanisms, though. Therefore, the observation of a particular scaling relation in the quantum regime could help to identify the dominant scattering mechanism.

In the case of impurity scattering treated in the SCBA, the first scaling relation is found by eliminating $g$ from Eqs.~(\ref{eq:sigmaxx-k2}) and (\ref{eq:RH-k2}), which gives the correct measure of the carrier density in the quantum regime:
	\begin{equation}
		\frac{6\pi}{5}\frac{\sigma_{xx}/(e^2/h)}{-|e|\RH}=n.
	\end{equation}
Conversely, by eliminating $n$, one finds a proper measure of the coupling constant,
	\begin{equation}
		\frac{1}{-|e|\RH\sqrt{\sigma_{xx}/(e^2/h)}}=\frac{1}{a^2}\frac{5}{2\sqrt{3}\pi^{5/2}}
		\left(\frac{g}{\mathrm{Hr}}\right)^2,
	\end{equation}
which is proportional to $n_c$ defined in Eq.~(\ref{eq:nc-k2}). For 2D materials in which the coupling constant could be tuned, for example by varying the concentration of impurities, these scaling laws should only hold in the quantum regime, which would provide another way to reveal the quantum to semiclassical crossover, beside the change of behavior in $\sigma_{xy}$ seen in Fig.~\ref{fig:Hall_SCBA_mu}(b). Concerning field-effect experiments, where the control parameter is $\mu$, another scaling relation may be constructed by combining Eqs.~(\ref{eq:sigmaxx-k2}), (\ref{eq:RH-k2}), and (\ref{eq:mu}),
	\begin{equation}
		\frac{\sqrt{\sigma_{xx}/(e^2/h)}}{-|e|\RH}=\frac{1}{a^2}\frac{5}{3\sqrt{3}\pi^{5/2}}
		\frac{\mu-\varepsilon_0}{\mathrm{Hr}},
	\end{equation}
which only depends on $g$ via the conduction threshold $\varepsilon_0$.

We point out that these scaling relations and the power laws obtained in the quantum regime are robust to changes in the system's dimensionality and may be observed as well in quasi-2D and 3D disordered metals. Indeed, Eqs.~(\ref{eq:A})--(\ref{eq:sigmaxy}), which define the conductivities, are valid in 3D for describing the transport in the plane perpendicular to the magnetic field. The power laws that we find at low density occur in 3D as well, namely $\sigma_{xx}\sim n^{2/3}$, $\sigma_{xy}\sim n$ versus density and $\sigma_{xx}\sim\mu-\varepsilon_0$, $\sigma_{xy}\sim(\mu-\varepsilon_0)^{3/2}$ versus chemical potential. To explain this, we note that the power laws can be traced back to three facts: (i) the conductivities given in Eqs.~(\ref{eq:sigmaxx_M_phi}) and (\ref{eq:sigmaxy_M_phi}) vary as $(\pi-\phi)^2$ and $(\pi-\phi)^3$ for $\phi\to\pi$; (ii) $\pi-\phi$ is proportional to $\mathrm{Im}\,\Sigma^{\mathrm{SCBA}}(\mu)\propto(\mu-\varepsilon_0)^{1/2}$; and (iii) the density varies as $(\mu-\varepsilon_0)^{3/2}$. The argument leading to $\mathrm{Im}\,\Sigma^{\mathrm{SCBA}}(\varepsilon)=-C(\mu-\varepsilon_0)^{1/2}$ does not use the specific form of $N_0(\varepsilon)$, as can be seen in Appendix~\ref{app:edge}. Therefore, this same square-root behavior occurs in 3D, fulfilling the condition (ii), although the dependences of the parameters $\varepsilon_0$ and $C$ on the coupling $g$ are different in 2D and 3D. The property (iii) follows from Eq.~(\ref{eq:SCBA_DOS}), which again does not depend on the specific form of $N_0(\varepsilon)$ and holds in 3D. Finally, the transport functions $\Phi_{xx}(E)$ and $\Phi_{xy}(E)$ vary as $E^{3/2}$ in 3D, such that Eqs.~(\ref{eq:sigmaxx_M_phi}) and (\ref{eq:sigmaxy_M_phi}) get replaced by different functions of $\phi$. These new functions nevertheless obey the conditions (i), as the exponent of the transport functions determines the coefficient, but not the leading exponent of the expansion in powers of $\pi-\phi$.

\begin{figure}[tb]
\includegraphics[width=\columnwidth]{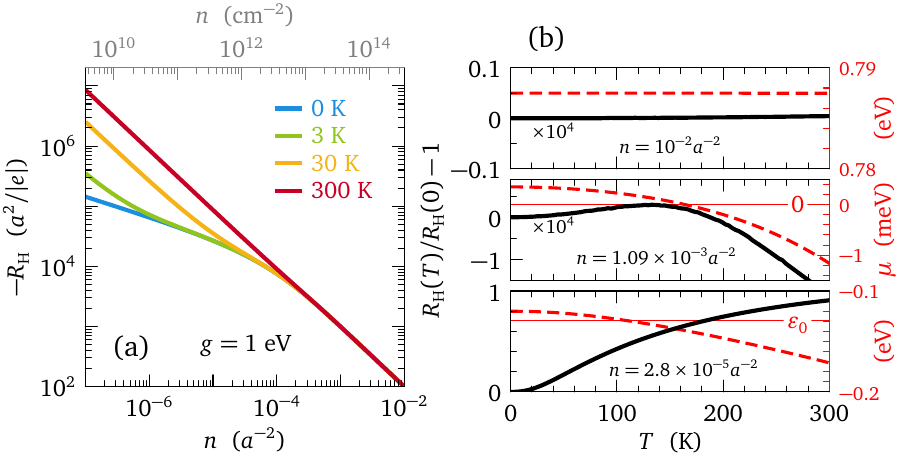}
\caption{\label{fig:T>0}
(a) Temperature dependence of the Hall constant for a parabolic band and the SCBA self-energy with $g=1$~eV and a cutoff $E_c=10$~Hr. (b) Hall constant and chemical potential vs $T$ for three electron densities. 
}
\end{figure}

Finally, one may wonder whether the temperature dependence could also provide some hint about the quantum regime. Indeed, an anomaly occurs in the temperature dependence of the Hall constant in the constant-scattering rate approximation, when the chemical potential traverses the edge of the noninteracting band \cite{Morpurgo_Giamarchi_2024_Hall_effect_low_density}. The anomaly is largest at the lowest densities. In the case of the SCBA, however, we only find a very weak anomaly. Figure~\ref{fig:T>0}(a) shows that $\RH$ returns to the semiclassical power law as the temperature is raised. The details of the temperature dependence are displayed in Fig.~\ref{fig:T>0}(b) for three representative carrier densities. For $n=10^{-2}a^{-2}$, $\RH$ has virtually no temperature dependence, as expected in the semiclassical regime. The chemical potential remains far from both the noninteracting and impurity band edges. For $n=2.8\times10^{-5}a^{-2}$, the chemical potential is just above the edge of the impurity band at $T=0$ and crosses it as $T$ increases. $\RH$ raises nonetheless monotonously towards $\RH^0$. At the density $1.09\times10^{-3}a^{-2}$, $\mu$ is just above the noninteracting band edge at $T=0$ and crosses it as $T$ increases. Here, we see a non-monotonous evolution of $\RH$ with increasing $T$, but the amplitude of the effect is tiny.

Thus, the large amplitude of the anomalous temperature dependence found in Ref.~\cite{Morpurgo_Giamarchi_2024_Hall_effect_low_density} appears to be peculiar to the constant-scattering rate approximation. In the SCBA, despite the absence of a strong anomaly, the temperature dependence evolves from being very weak in the semiclassical regime to being pronounced deep in the quantum regime, where $\RH$ increases by a factor $2$ between base and room temperatures with the parameters of Fig.~\ref{fig:T>0}.

\subsection{Anderson localization and validity of the SCBA}

\begin{figure}[tb]
\includegraphics[width=0.9\columnwidth]{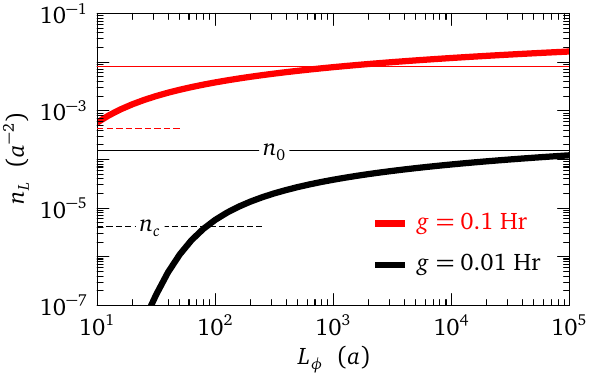}
\caption{\label{fig:localization}
Estimated density $n_L$ of the metal-insulator transition due to Anderson localization vs electron phase-coherence length $L_{\phi}$ for two values of the coupling $g$. $n_0$ (solid horizontal lines) and $n_c$ (dashed horizontal lines) correspond to the arrows in Fig.~\ref{fig:k2-SCBA}(d).
}
\end{figure}

In the absence of inelastic scattering, a two-dimensional electron gas subject to elastic disorder is expected to become insulating if the localization length $\xi\sim\ell e^{\frac{\pi}{2}k_{\mathrm{F}}\ell}$ is shorter than the typical sample size $L$ \cite{Anderson_1958_localisation, Review_Lee_Ramakrishnan_1985_disorder}. Here $k_{\mathrm{F}}$ is the Fermi wavevector and $\ell$ is the elastic mean free path. The localization is unlikely at high density $k_{\mathrm{F}}\ell\gg1$, given the exponential increase of $\xi$, but it is expected if $k_{\mathrm{F}}\ell\lesssim1$. In real materials, inelastic scattering is always present and it reduces the electron phase-coherence length $L_{\phi}$ to values smaller than the sample size. This renormalizes the metal-insulator transition to a lower density obeying $\xi=L_{\phi}$. On the other hand, if $L_{\phi}\ll L$, self-averaging is expected to occur for uncorrelated disorder, such that the effect of disorder can be captured via the impurity-average technique. The SCBA should therefore be reliable if $L_{\phi}\ll L$ and if the disorder is weak (see Sec.~\ref{sec:SCBA}). Evaluating the $n$-dependent mean free path as $\ell=v_{\mathrm{F}}\tau$ with $v_{\mathrm{F}}$ the Fermi velocity and $\tau$ the quasiparticle lifetime, which in the SCBA is given by $\tau=\hbar/[2|\mathrm{Im}\,\Sigma^{\mathrm{SCBA}}(\mu)|]$, and solving the equation $\xi(n_L)=L_{\phi}$, we extract the localization density $n_L$ plotted in Fig.~\ref{fig:localization}.

For the coupling $g=0.01$~Hr, the carrier density at which the metal-insulator transition is expected to occur is lower than $n_0$, provided that $L_{\phi}$ is shorter than $\sim10^{5}a$, i.e., a few micrometers for a band mass equal to the electron mass. Therefore, the possibly to observe deviations from the semiclassical behavior is not seriously limited by the localization phenomenon. The quantum regime is fully reached at a significantly lower density $n_c$ [see Fig.~\ref{fig:k2-SCBA}(b)], where carriers are expected to be localized unless $L_{\phi}<10^2a$, which would be on the scale of nanometers. At the point where $n_L=n_c$, the mean free path $\ell\sim54a$ is still significantly shorter than $L_{\phi}$. Estimating the inelastic scattering time via $L_{\phi}=v_{\mathrm{F}}\sqrt{\tau\tau_{\mathrm{inel}}}$ \cite{Review_Lee_Ramakrishnan_1985_disorder}, we find that $\tau_{\mathrm{inel}}\sim 600$~fs when $n_L=n_c$ for a mass equal to the electron mass, which is still much longer than typical values in metals. For the stronger coupling $g=0.1$~Hr, the quantum regime is presumably out of reach. We therefore expect that the quantum regime may be reached before the carriers localize in systems with short inelastic scattering time and sufficiently weak impurity scattering, and that our results are relevant in this regime if $L_{\phi}\ll L$.

\section{Conclusion}
\label{sec:conclusion}

We have introduced a microscopic model describing electrons interacting with weak pointlike disorder in a magnetic field, and we solved it for the linear conductivity tensor and the Hall constant in two dimensions. Analytical results were derived at zero temperature and low carrier density, while the other regimes were addressed by accurate numerics using a novel methodology based on piecewise functions. Like in a model studied previously with a phenomenological constant scattering rate, we find a crossover from a semiclassical regime at high density, where Drude-like conductivities are observed, to a low-density quantum regime where they have different functional forms, which at zero temperature are power laws of the density and disorder strength. Unlike in the phenomenological model, however, these power laws are robust to variations of the electron dispersion and change of the spatial dimension. The power laws allow one to define functions of the longitudinal and transverse conductivities that provide direct measures of the carrier density and scattering strength, similar to what the Hall carrier density and the mobility do in the semiclassical regime. We argued that the quantum regime may be within reach of the state-of-the-art field-effect experiments on gated semiconductors.

Our previous study \cite{Morpurgo_Giamarchi_2024_Hall_effect_low_density} has shown that anisotropy in the electron dispersion enhances the parameter range covered by the quantum regime. Although we have not addressed this question here, we expect a similar effect in the present model. Indeed, we find that for an anisotropic parabolic dispersion with $m_{yy}>m_{xx}$, the crossover density $n_c$ increases by a factor $m_{yy}/m_{xx}$. The power laws found in this study ultimately originate in the fact that the scattering rate and the interacting DOS vanish as square roots at the bottom of the impurity band. Different local scattering models may lead to different behaviors at the interacting band edge and thus to different power laws. The case of a local Fermi liquid would be worth investigating. This is more difficult than the SCBA treated here, because the Fermi-liquid self-energy develops around the chemical potential and therefore depends in a nontrivial way on the density, while the SCBA self-energy does not. Longer-term perspectives include a multiband generalization of local scattering models, which would enable accurate solutions at low density for systems like graphene. Much more challenging is the generalization to extended disorder or other nonlocal scattering models, for which the vertex corrections cannot be ignored: how to deal accurately with those problems in the low-density limit remains an open question.

\begin{acknowledgments}
We are thankful to M. Melegari, A. Morpurgo and N. Prokofiev for fruitful discussions. This work was supported by the Swiss National Science Foundation under Division II (Grant No. 200020-219400). 
\end{acknowledgments}

\section{Data Availability}

The data that support the findings of this article and the related computer codes are openly available \cite{yareta}.

\appendix

\section{Fast integral transforms using piecewise functions}
\label{app:piecewise}

Here, using Eq.~(\ref{eq:SCBA}) as an example, we illustrate how the use of piecewise functions enables fast numerical integral transforms. The right-hand side of Eq.~(\ref{eq:SCBA}) is of the form
	\begin{equation}\label{eq:Hilbert}
		\int_{-\infty}^{\infty}dx\,\frac{f(x)}{z-x},
	\end{equation}
where the function $f(x)$ plays the role of $(ga)^2N_0(x)$ and the complex number $z\in\mathbb{C}\setminus\mathbb{R}$ plays the role of $\varepsilon-\Sigma^{\mathrm{SCBA}}(\varepsilon)$. Equation~(\ref{eq:Hilbert}) is an example of the integral transform of the function $f(x)$, which uses the kernel $\mathcal{K}(x,z)=1/(z-x)$ and is a generalization of the Hilbert transform. Equations~(\ref{eq:N}), (\ref{eq:sigmaxx}), and (\ref{eq:sigmaxy}) also involve integral transforms of the functions $N_0(E)$, $\Phi_{xx}(E)$, and $\Phi_{xy}(E)$ using other kernels. Our numerical approach for evaluating these integral transforms is to represent the function to be transformed, e.g., $f(x)$ in Eq.~(\ref{eq:Hilbert}), in a piecewise manner, using in each piece mathematical functions $F_i(x)$ that are sufficiently flexible to capture the behavior of $f(x)$ accurately, and sufficiently simple for the solution of the equation $\frac{d}{dx}\mathcal{F}_i(x,z)=F_i(x)\mathcal{K}(x,z)$ to be known. Thus, the calculation of the integral transform reduces to evaluating $\mathcal{F}_i(x,z)$ at the boundaries of each piece. The approach is useful if the number of pieces is much smaller than the number of function evaluations that would be required to calculate the integral transform using a quadrature. Hence the choice of the functions $F_i(x)$ is crucial. As the functions like $N_0(E)$ that must be piecewise-approximated can present singularities and nonanalyticities, the approach may fail if $F_i(x)$ are just polynomial functions---like for instance cubic splines---that cannot capture these nonanalytic behaviors accurately.

\begin{figure}[tb]
\includegraphics[width=\columnwidth]{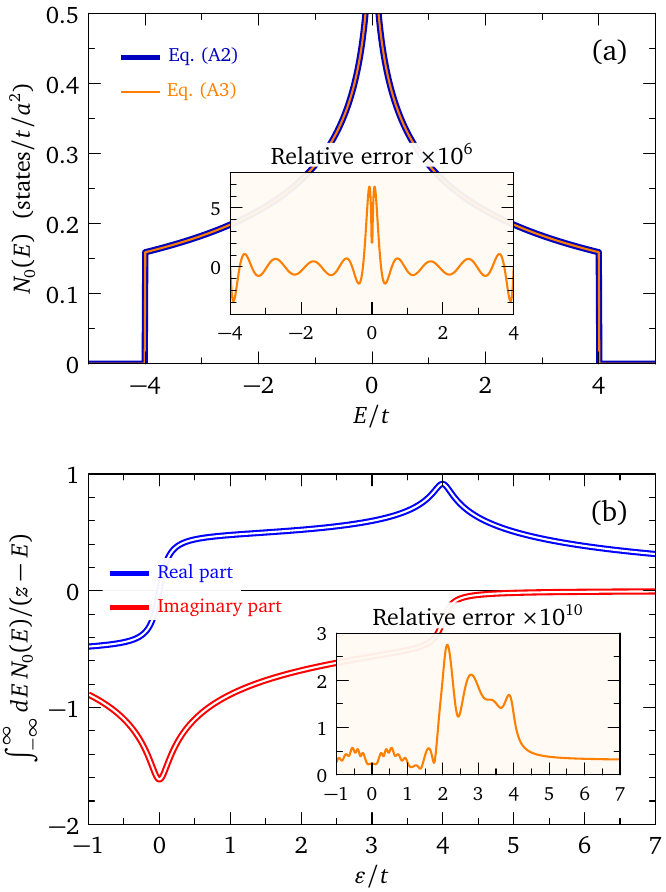}
\caption{\label{fig:piecewise}
(a) Approximation of the square-lattice tight-binding DOS (dark blue) by elementary mathematical functions (orange). Inset: Relative error of the approximation ($<10^{-5}$). (b) Integral transform of the DOS vs $z=\varepsilon+i\delta$ for $\delta=0.1t$. Thick lines are obtained with a quadrature using $N_0(E)$ given by Eq.~(\ref{eq:N0_cosk}); thin white lines use a piecewise approximation with $<10^{-8}$ relative accuracy. Inset: Relative error on the integral transform ($<10^{-9}$). 
}
\end{figure}

The DOS of the square-lattice nearest-neighbor tight-binding model is
	\begin{equation}\label{eq:N0_cosk}
		N_0(E) = \frac{1}{\pi^2a^2t}K\left(1-\left(\frac{E}{4t}\right)^2\right)\theta(4t-|E|),
	\end{equation}
where $K$ is the complete elliptic integral of the first kind. As it turns out, this function can be represented with $<10^{-5}$ relative accuracy as a \emph{single} piece,
	\begin{equation}\label{eq:N0_piecewise}
		N_0(E)\approx\left(a_0\ln|E|+\sum_{k=1}^{11}a_k|E|^{k-1}\right)\theta(4t-|E|)
	\end{equation}
with the following numerical parameters:
\begin{align*}
a_0 &= \texttt{-1.013211836423378e-01}\\[-0.2em]
a_1 &= \texttt{ 2.809219621590210e-01}\\[-0.2em]
a_2 &= \texttt{ 2.049497230383165e-04}\\[-0.2em]
a_3 &= \texttt{ 5.167691304757171e-03}\\[-0.2em]
a_4 &= \texttt{-4.729671436113094e-03}\\[-0.2em]
a_5 &= \texttt{ 3.854466083613651e-03}\\[-0.2em]
a_6 &= \texttt{-2.371290515504572e-03}\\[-0.2em]
a_7 &= \texttt{ 1.017346738868159e-03}\\[-0.2em]
a_8 &= \texttt{-2.919228439402373e-04}\\[-0.2em]
a_9 &= \texttt{ 5.317878986488775e-05}\\[-0.2em]
a_{10} &= \texttt{-5.547047143635150e-06}\\[-0.2em]
a_{11} &= \texttt{ 2.519224597696242e-07}
\end{align*}
We use a better approximation with more pieces and $<10^{-8}$ relative accuracy in our calculations of the conductivities. Figure~\ref{fig:piecewise}(a) shows a comparison of Eqs.~(\ref{eq:N0_cosk}) and (\ref{eq:N0_piecewise}). The evaluation of the integral in Eq.~(\ref{eq:Hilbert}) is straightforward for a function like Eq.~(\ref{eq:N0_piecewise}), because the differential equation
	\begin{equation}
		\frac{d}{dx}\mathcal{F}(x,z)=\left(a_0\ln x+\sum_{k=1}^{11}a_kx^{k-1}\right)\frac{1}{z-x}
	\end{equation}
admits a closed solution in terms of the second-order polylogarithm $\mathrm{Li}_2$ and the hypergeometric function ${_2}F_1$:
	\begin{multline}
		\mathcal{F}(x,z)=-a_0\left[\ln(x)\ln\left(1-\frac{x}{z}\right)
		+\mathrm{Li}_2\left(\frac{x}{z}\right)\right]\\
		+\frac{1}{z}\sum_{k=1}^{11}\frac{a_kx^k}{k}{_2}F_1\left(1,k,k+1,\frac{x}{z}\right).
	\end{multline}
It follows that the integral transform is reduced to a small number of calls to these special functions:
	\begin{multline}
		\int_{-\infty}^{\infty}dE\,\frac{N_0(E)}{z-E}\approx\mathcal{F}(4t,z)-\mathcal{F}(0^+,z)\\
		-\mathcal{F}(4t,-z)+\mathcal{F}(0^+,-z),
	\end{multline}
where use has been made of the property $N_0(E)=N_0(-E)$. Figure~\ref{fig:piecewise}(b) shows a comparison of the integral transform performed by quadratures and without quadrature, using our piecewise approximation of $N_0(E)$ with $<10^{-8}$ accuracy. The relative error on the integral transform is smaller than the relative error on the DOS.

\section{Accuracy of the SCBA for dense weak disorder}
\label{app:ED}

To assess the accuracy of the SCBA in the regime of dense and weak disorder, we performed exact-diagonalization (ED) for the square-lattice model using an $N\times N$ lattice ($N=100$) with periodic boundary conditions and an uncorrelated disorder potential on every site, uniformly distributed within $\pm\sqrt{6}g$. This model of disorder is different from the identical impurities discussed in the introductory part of Sec.~\ref{sec:SCBA}, although both models correspond to the same SCBA given by Eq.~(\ref{eq:SCBA}). We computed the spatially averaged Green's function $G(\vec{k},\varepsilon)=(1/N^2)\sum_{\vec{r}\vec{r}'}e^{-i\vec{k}\cdot(\vec{r}-\vec{r}')}G(\vec{r},\vec{r}',\varepsilon)$ and averaged it over $10^4$ configurations of the disorder. The Green's function was evaluated as $G(\vec{r},\vec{r}',\varepsilon)=\sum_n\varphi^*_n(\vec{r})\varphi^{}_n(\vec{r}')/(\varepsilon-E_n+i\delta)$, where $E_n$ and $\varphi_n(\vec{r})$ are the eigenvalues and eigenfunctions, respectively, and the energy-broadening parameter $\delta=64t/N^2$ is of the order of the typical interlevel spacing. From the disorder-averaged Green's function, we deduced the self-energy as $\Sigma(\vec{k},\varepsilon)=\varepsilon+i\delta-E_{\vec{k}}-1/G(\vec{k},\varepsilon)$.

\begin{figure}[tb]
\includegraphics[width=\columnwidth]{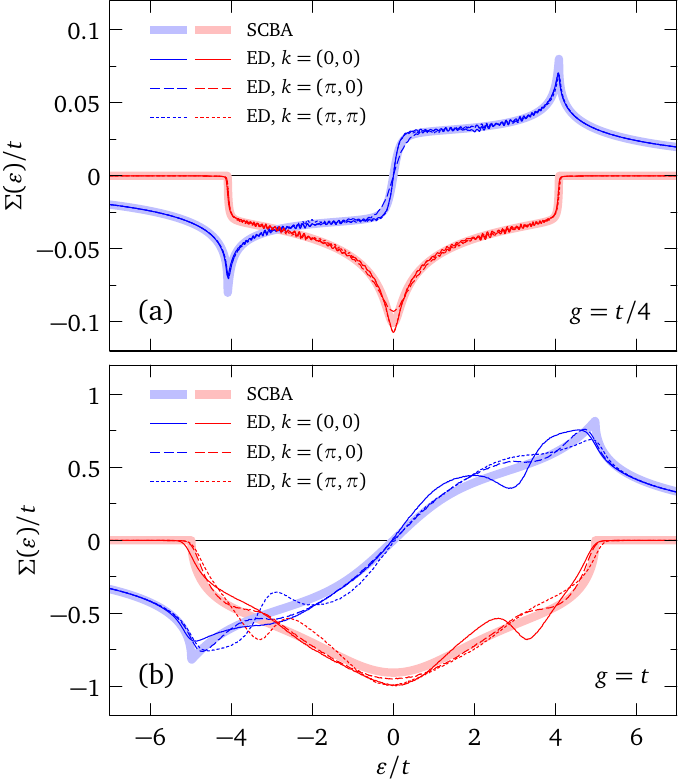}
\caption{\label{fig:ED}
Real part (blue) and imaginary part (red) of the disorder-averaged self-energy obtained using the momentum-independent self-consistent Born approximation (SCBA, thick lines) and exact diagonalization for three representative momenta (ED, thin lines) for (a) weak and (b) strong disorder potential.
}
\end{figure}

In Fig.~\ref{fig:ED}(a), we compare the SCBA with the ED results for a coupling $g=t/4$, which is larger than all the values used to calculate the conductivities in Fig.~\ref{fig:cosk-SCBA}. It is seen that the SCBA is accurate in that case. Within the uncertainties associated with the finite size of the lattice and the finite number of disorder realizations, the ED self-energy shows very little momentum dependence and an energy dependence that is hardly distinguishable from the SCBA. In particular, the edge $\varepsilon_0$ and the amplitude $C$ of the square-root behavior of $\mathrm{Im}\,\Sigma$ near the edge, which are the key ingredients that determine the asymptotic low-density behavior of the conductivities, appear to be accurately captured by the SCBA. Figure~\ref{fig:ED}(b) illustrates the deviations from the SCBA that develop at stronger coupling. The ED self-energy acquires a weak momentum dependence and breaks particle-hole symmetry at a given $\vec{k}$ (although the momentum-averaged self-energy remains particle-hole symmetric). Despite these deviations, the overall shape and magnitude of the self-energy are still very well described by the SCBA.

\section{SCBA self-energy near the band edge}
\label{app:edge}

The imaginary part of the SCBA self-energy given by Eq.~(\ref{eq:SCBA}) vanishes at a model-dependent [i.e., $N_0(E)$- and $g$-dependent] energy $\varepsilon_0$ that is below and above the noninteracting band edges. We focus on the lower edge for definiteness and investigate the behavior of the self-energy in the vicinity of $\varepsilon_0$. With the notation $\Sigma^{\mathrm{SCBA}}(\varepsilon)=\Sigma_1(\varepsilon)+i\Sigma_2(\varepsilon)$, we have
	\begin{subequations}\begin{align}
		\label{eq:Sigma1}
    	\Sigma_1(\varepsilon)&=(ga)^2\int_{-\infty}^{\infty}dE\!
        \,\frac{N_0(E)[\varepsilon-\Sigma_1(\varepsilon)-E]}
    	{[\varepsilon-\Sigma_1(\varepsilon)-E]^2+[\Sigma_2(\varepsilon)]^2}\\
		\label{eq:Sigma2}
    	\Sigma_2(\varepsilon)&=(ga)^2\int_{-\infty}^{\infty}dE\,\!
        \frac{N_0(E)\Sigma_2(\varepsilon)}
    	{[\varepsilon-\Sigma_1(\varepsilon)-E]^2+[\Sigma_2(\varepsilon)]^2},
	\end{align}\end{subequations}
where it is seen that a factor $\Sigma_2(\varepsilon)$ cancels in the second relation. Setting $\Sigma_2(\varepsilon_0)=0$ and developing this second relation around $\varepsilon=\varepsilon_0$ gives
	\begin{multline}\label{eq:expansion1}
		\frac{1}{(ga)^2}=\int_{-\infty}^{\infty}dE\,\frac{N_0(E)}{[\varepsilon_0-\Sigma_1(\varepsilon_0)-E]^2}
		-2(\varepsilon-\varepsilon_0)\\ \times[1-\Sigma_1'(\varepsilon_0)]
		\int_{-\infty}^{\infty}dE\,\frac{N_0(E)}{[\varepsilon_0-\Sigma_1(\varepsilon_0)-E]^3}+\ldots
	\end{multline}
This development is valid if $\Sigma_1'(\varepsilon_0)$ is finite, which is only true when $\varepsilon$ approaches $\varepsilon_0$ from above. On the other hand, developing Eq.~(\ref{eq:Sigma2}) around $\Sigma_2(\varepsilon)=0$, we obtain
	\begin{multline}\label{eq:expansion2}
		\frac{1}{(ga)^2}=\int_{-\infty}^{\infty}dE\,\frac{N_0(E)}{[\varepsilon-\Sigma_1(\varepsilon)-E]^2}\\
		-[\Sigma_2(\varepsilon)]^2
		\int_{-\infty}^{\infty}dE\,\frac{N_0(E)}{[\varepsilon-\Sigma_1(\varepsilon)-E]^4}+\ldots
	\end{multline}
Again, Eq.~(\ref{eq:expansion2}) is only valid for $\varepsilon>\varepsilon_0$, where $\Sigma_2(\varepsilon)$ approaches zero continuously. Evaluating Eq.~(\ref{eq:expansion2}) at leading order in $\varepsilon-\varepsilon_0$ and combining with Eq.~(\ref{eq:expansion1}), we find that $\Sigma_2(\varepsilon)$ varies as a square root close to $\varepsilon_0$:
	\begin{equation*}
		[\Sigma_2(\varepsilon)]^2=2(\varepsilon-\varepsilon_0)[1-\Sigma_1'(\varepsilon_0)]
		\frac{\int_{-\infty}^{\infty}dE\,\frac{N_0(E)}{[\varepsilon_0-\Sigma_1(\varepsilon_0)-E]^3}}
		{\int_{-\infty}^{\infty}dE\,\frac{N_0(E)}{[\varepsilon_0-\Sigma_1(\varepsilon_0)-E]^4}}.
	\end{equation*}
Thus $\Sigma_2(\varepsilon)=-C\sqrt{\varepsilon-\varepsilon_0}$ close to a lower band edge, with a model-dependent constant $C$ given by
	\begin{equation}\label{eq:C}
		C=\sqrt{2[1-\Sigma_1'(\varepsilon_0)]
		\frac{\int_{-\infty}^{\infty}dE\,\frac{N_0(E)}{[\varepsilon_0-\Sigma_1(\varepsilon_0)-E]^3}}
		{\int_{-\infty}^{\infty}dE\,\frac{N_0(E)}{[\varepsilon_0-\Sigma_1(\varepsilon_0)-E]^4}}}.
	\end{equation}
The derivative $\Sigma_1'(\varepsilon)$ is discontinuous at the interacting band edge $\varepsilon_0$ (see below) so the value to be used in Eq.~(\ref{eq:C}) is the value inside the band.

Below the band edge, where $\Sigma_2(\varepsilon)=0$, the derivative $\Sigma_1'(\varepsilon)$ may be obtained from Eq.~(\ref{eq:Sigma1}):
	\begin{equation*}
		\Sigma_1'(\varepsilon<\varepsilon_0)=\frac{1}{1-\left\{(ga)^2\int_{-\infty}^{\infty}dE\,\frac{N_0(E)}
    	{[\varepsilon-\Sigma_1(\varepsilon)-E]^2}\right\}^{-1}}.
	\end{equation*}
Furthermore, evaluating Eq.~(\ref{eq:Sigma2}) at $\varepsilon_0$ with $\Sigma_2(\varepsilon_0)=0$ gives the relation
	\begin{equation*}
		1=(ga)^2\int_{-\infty}^{\infty}dE\,\frac{N_0(E)}{[\varepsilon_0-\Sigma_1(\varepsilon_0)-E]^2}.
	\end{equation*}
These two expressions show that $\Sigma_1'(\varepsilon)$ diverges when $\varepsilon$ approaches $\varepsilon_0$ from below, as the quantity in curly braces approaches $1$.

For the parabolic dispersion, the implicit relation Eq.~(\ref{eq:SCBA_parabolic}) allows one to extract closed expressions for $\varepsilon_0$ and $C$. We start by expanding Eq.~(\ref{eq:SCBA_parabolic}) in power of $\Sigma_2(\varepsilon)$, as appropriate for $\varepsilon\gtrsim\varepsilon_0$, up to second order. The imaginary part of this expansion yields a second-order algebraic equation for $\Sigma_1(\varepsilon)$, whose solution is
	\begin{equation}\label{eq:Sigma1_eps0}
		\Sigma_1(\varepsilon)=\varepsilon+\frac{E_c}{2}\left(\sqrt{1+\gamma}-1\right),
	\end{equation}
where $\gamma=4(ga)^2N_0/E_c$. This shows that $\Sigma_1(\varepsilon)$ increases linearly for $\varepsilon\gtrsim\varepsilon_0$, as seen in Figs.~\ref{fig:SCBA_k2} and \ref{fig:SCBA_cosk}. As the self-energy is continuous at $\varepsilon_0$, we can deduce expressions for $\Sigma_1(\varepsilon_0)$ and $\varepsilon_0-\Sigma_1(\varepsilon_0)$, which we insert back into Eq.~(\ref{eq:SCBA_parabolic}), together with $\Sigma_2(\varepsilon_0)=0$, to get
	\begin{multline*}
		\varepsilon_0+\frac{E_c}{2}\left(\sqrt{1+\gamma}-1\right)=(ga)^2N_0\left\{
		\ln\left[\frac{E_c}{2}\left(\sqrt{1+\gamma}-1\right)\right]\right.\\
		\left.-\ln\left[\frac{E_c}{2}\left(\sqrt{1+\gamma}-1\right)+E_c\right]\right\}.
	\end{multline*}
Solving for $\varepsilon_0$, we obtain Eq.~(\ref{eq:eps0_parabolic}). On the other hand, the real part of the expansion of Eq.~(\ref{eq:SCBA_parabolic}), complemented with the expression of $\varepsilon_0$, yields a equation for $[\Sigma_2(\varepsilon)]^2$, whose solution is
	\begin{equation}\label{eq:Sigma2_eps0}
		\Sigma_2(\varepsilon)=-\frac{\sqrt{E_c\gamma/2}}{(1+\gamma)^{1/4}}\sqrt{\varepsilon-\varepsilon_0},
	\end{equation}
which leads to Eq.~(\ref{eq:C_parabolic}).


\begin{thebibliography}{56}%
\makeatletter
\providecommand \@ifxundefined [1]{%
 \@ifx{#1\undefined}
}%
\providecommand \@ifnum [1]{%
 \ifnum #1\expandafter \@firstoftwo
 \else \expandafter \@secondoftwo
 \fi
}%
\providecommand \@ifx [1]{%
 \ifx #1\expandafter \@firstoftwo
 \else \expandafter \@secondoftwo
 \fi
}%
\providecommand \natexlab [1]{#1}%
\providecommand \enquote  [1]{``#1''}%
\providecommand \bibnamefont  [1]{#1}%
\providecommand \bibfnamefont [1]{#1}%
\providecommand \citenamefont [1]{#1}%
\providecommand \href@noop [0]{\@secondoftwo}%
\providecommand \href [0]{\begingroup \@sanitize@url \@href}%
\providecommand \@href[1]{\@@startlink{#1}\@@href}%
\providecommand \@@href[1]{\endgroup#1\@@endlink}%
\providecommand \@sanitize@url [0]{\catcode `\\12\catcode `\$12\catcode
  `\&12\catcode `\#12\catcode `\^12\catcode `\_12\catcode `\%12\relax}%
\providecommand \@@startlink[1]{}%
\providecommand \@@endlink[0]{}%
\providecommand \url  [0]{\begingroup\@sanitize@url \@url }%
\providecommand \@url [1]{\endgroup\@href {#1}{\urlprefix }}%
\providecommand \urlprefix  [0]{URL }%
\providecommand \Eprint [0]{\href }%
\providecommand \doibase [0]{https://doi.org/}%
\providecommand \selectlanguage [0]{\@gobble}%
\providecommand \bibinfo  [0]{\@secondoftwo}%
\providecommand \bibfield  [0]{\@secondoftwo}%
\providecommand \translation [1]{[#1]}%
\providecommand \BibitemOpen [0]{}%
\providecommand \bibitemStop [0]{}%
\providecommand \bibitemNoStop [0]{.\EOS\space}%
\providecommand \EOS [0]{\spacefactor3000\relax}%
\providecommand \BibitemShut  [1]{\csname bibitem#1\endcsname}%
\let\auto@bib@innerbib\@empty
\bibitem [{\citenamefont {Ando}\ \emph {et~al.}(1982)\citenamefont {Ando},
  \citenamefont {Fowler},\ and\ \citenamefont
  {Stern}}]{Review_1982_Ando_2d_semiconductors}%
  \BibitemOpen
  \bibfield  {author} {\bibinfo {author} {\bibfnamefont {T.}~\bibnamefont
  {Ando}}, \bibinfo {author} {\bibfnamefont {A.~B.}\ \bibnamefont {Fowler}},\
  and\ \bibinfo {author} {\bibfnamefont {F.}~\bibnamefont {Stern}},\ }\bibfield
   {title} {\bibinfo {title} {Electronic properties of two-dimensional
  systems},\ }\href {https://doi.org/10.1103/RevModPhys.54.437} {\bibfield
  {journal} {\bibinfo  {journal} {Rev. Mod. Phys.}\ }\textbf {\bibinfo {volume}
  {54}},\ \bibinfo {pages} {437} (\bibinfo {year} {1982})}\BibitemShut
  {NoStop}%
\bibitem [{\citenamefont {Tsumura}\ \emph {et~al.}(1986)\citenamefont
  {Tsumura}, \citenamefont {Koezuka},\ and\ \citenamefont
  {Ando}}]{Tsumura_Ando_1986_pionnier_field_effect_transistor}%
  \BibitemOpen
  \bibfield  {author} {\bibinfo {author} {\bibfnamefont {A.}~\bibnamefont
  {Tsumura}}, \bibinfo {author} {\bibfnamefont {H.}~\bibnamefont {Koezuka}},\
  and\ \bibinfo {author} {\bibfnamefont {T.}~\bibnamefont {Ando}},\ }\bibfield
  {title} {\bibinfo {title} {Macromolecular electronic device: Field‐effect
  transistor with a polythiophene thin film},\ }\href
  {https://doi.org/10.1063/1.97417} {\bibfield  {journal} {\bibinfo  {journal}
  {Appl. Phys. Lett.}\ }\textbf {\bibinfo {volume} {49}},\ \bibinfo {pages}
  {1210} (\bibinfo {year} {1986})}\BibitemShut {NoStop}%
\bibitem [{\citenamefont {Novoselov}\ \emph {et~al.}(2004)\citenamefont
  {Novoselov}, \citenamefont {Geim}, \citenamefont {Morozov}, \citenamefont
  {Jiang}, \citenamefont {Zhang}, \citenamefont {Dubonos}, \citenamefont
  {Grigorieva},\ and\ \citenamefont
  {Firsov}}]{Novoselov_Firsov_2004_graphene_field_effect}%
  \BibitemOpen
  \bibfield  {author} {\bibinfo {author} {\bibfnamefont {K.~S.}\ \bibnamefont
  {Novoselov}}, \bibinfo {author} {\bibfnamefont {A.~K.}\ \bibnamefont {Geim}},
  \bibinfo {author} {\bibfnamefont {S.~V.}\ \bibnamefont {Morozov}}, \bibinfo
  {author} {\bibfnamefont {D.}~\bibnamefont {Jiang}}, \bibinfo {author}
  {\bibfnamefont {Y.}~\bibnamefont {Zhang}}, \bibinfo {author} {\bibfnamefont
  {S.~V.}\ \bibnamefont {Dubonos}}, \bibinfo {author} {\bibfnamefont {I.~V.}\
  \bibnamefont {Grigorieva}},\ and\ \bibinfo {author} {\bibfnamefont {A.~A.}\
  \bibnamefont {Firsov}},\ }\bibfield  {title} {\bibinfo {title} {Electric
  field effect in atomically thin carbon films},\ }\href
  {https://doi.org/10.1126/science.1102896} {\bibfield  {journal} {\bibinfo
  {journal} {Science}\ }\textbf {\bibinfo {volume} {306}},\ \bibinfo {pages}
  {666} (\bibinfo {year} {2004})}\BibitemShut {NoStop}%
\bibitem [{\citenamefont {Manzeli}\ \emph {et~al.}(2017)\citenamefont
  {Manzeli}, \citenamefont {Ovchinnikov}, \citenamefont {Pasquier},
  \citenamefont {Yazyev},\ and\ \citenamefont
  {Kis}}]{Review_Manzeli_Kis_2017_TMD}%
  \BibitemOpen
  \bibfield  {author} {\bibinfo {author} {\bibfnamefont {S.}~\bibnamefont
  {Manzeli}}, \bibinfo {author} {\bibfnamefont {D.}~\bibnamefont
  {Ovchinnikov}}, \bibinfo {author} {\bibfnamefont {D.}~\bibnamefont
  {Pasquier}}, \bibinfo {author} {\bibfnamefont {O.~V.}\ \bibnamefont
  {Yazyev}},\ and\ \bibinfo {author} {\bibfnamefont {A.}~\bibnamefont {Kis}},\
  }\bibfield  {title} {\bibinfo {title} {{2D} transition metal
  dichalcogenides},\ }\href {https://doi.org/10.1038/natrevmats.2017.33}
  {\bibfield  {journal} {\bibinfo  {journal} {Nature Reviews Materials}\
  }\textbf {\bibinfo {volume} {2}},\ \bibinfo {pages} {17033} (\bibinfo {year}
  {2017})}\BibitemShut {NoStop}%
\bibitem [{\citenamefont {Cao}\ \emph {et~al.}(2018)\citenamefont {Cao},
  \citenamefont {Fatemi}, \citenamefont {Fang}, \citenamefont {Watanabe},
  \citenamefont {Taniguchi}, \citenamefont {Kaxiras},\ and\ \citenamefont
  {Jarillo-Herrero}}]{Cao_Jarillo_Herrero_2018_bilayer_graphene_magic_angle}%
  \BibitemOpen
  \bibfield  {author} {\bibinfo {author} {\bibfnamefont {Y.}~\bibnamefont
  {Cao}}, \bibinfo {author} {\bibfnamefont {V.}~\bibnamefont {Fatemi}},
  \bibinfo {author} {\bibfnamefont {S.}~\bibnamefont {Fang}}, \bibinfo {author}
  {\bibfnamefont {K.}~\bibnamefont {Watanabe}}, \bibinfo {author}
  {\bibfnamefont {T.}~\bibnamefont {Taniguchi}}, \bibinfo {author}
  {\bibfnamefont {E.}~\bibnamefont {Kaxiras}},\ and\ \bibinfo {author}
  {\bibfnamefont {P.}~\bibnamefont {Jarillo-Herrero}},\ }\bibfield  {title}
  {\bibinfo {title} {Unconventional superconductivity in magic-angle graphene
  superlattices},\ }\href {https://doi.org/10.1038/nature26160} {\bibfield
  {journal} {\bibinfo  {journal} {Nature}\ }\textbf {\bibinfo {volume} {556}},\
  \bibinfo {pages} {43} (\bibinfo {year} {2018})}\BibitemShut {NoStop}%
\bibitem [{\citenamefont {Andrei}\ \emph {et~al.}(2021)\citenamefont {Andrei},
  \citenamefont {Efetov}, \citenamefont {Jarillo-Herrero}, \citenamefont
  {MacDonald}, \citenamefont {Mak}, \citenamefont {Senthil}, \citenamefont
  {Tutuc}, \citenamefont {Yazdani},\ and\ \citenamefont
  {Young}}]{Review_Andrei_Young_2021_Moire_materials}%
  \BibitemOpen
  \bibfield  {author} {\bibinfo {author} {\bibfnamefont {E.~Y.}\ \bibnamefont
  {Andrei}}, \bibinfo {author} {\bibfnamefont {D.~K.}\ \bibnamefont {Efetov}},
  \bibinfo {author} {\bibfnamefont {P.}~\bibnamefont {Jarillo-Herrero}},
  \bibinfo {author} {\bibfnamefont {A.~H.}\ \bibnamefont {MacDonald}}, \bibinfo
  {author} {\bibfnamefont {K.~F.}\ \bibnamefont {Mak}}, \bibinfo {author}
  {\bibfnamefont {T.}~\bibnamefont {Senthil}}, \bibinfo {author} {\bibfnamefont
  {E.}~\bibnamefont {Tutuc}}, \bibinfo {author} {\bibfnamefont
  {A.}~\bibnamefont {Yazdani}},\ and\ \bibinfo {author} {\bibfnamefont {A.~F.}\
  \bibnamefont {Young}},\ }\bibfield  {title} {\bibinfo {title} {The marvels of
  moir{\'e} materials},\ }\href {https://doi.org/10.1038/s41578-021-00284-1}
  {\bibfield  {journal} {\bibinfo  {journal} {Nat. Rev. Mat.}\ }\textbf
  {\bibinfo {volume} {6}},\ \bibinfo {pages} {201} (\bibinfo {year}
  {2021})}\BibitemShut {NoStop}%
\bibitem [{\citenamefont {Kopp}\ and\ \citenamefont
  {Mannhart}(2009)}]{Kopp_Mannhart_2009_compute_capacitance}%
  \BibitemOpen
  \bibfield  {author} {\bibinfo {author} {\bibfnamefont {T.}~\bibnamefont
  {Kopp}}\ and\ \bibinfo {author} {\bibfnamefont {J.}~\bibnamefont
  {Mannhart}},\ }\bibfield  {title} {\bibinfo {title} {Calculation of the
  capacitances of conductors: Perspectives for the optimization of electronic
  devices},\ }\href {https://doi.org/10.1063/1.3197246} {\bibfield  {journal}
  {\bibinfo  {journal} {J. Appl. Phys.}\ }\textbf {\bibinfo {volume} {106}},\
  \bibinfo {pages} {064504} (\bibinfo {year} {2009})}\BibitemShut {NoStop}%
\bibitem [{\citenamefont {Li}\ \emph {et~al.}(2011)\citenamefont {Li},
  \citenamefont {Nardes}, \citenamefont {Liang}, \citenamefont {Shaheen},
  \citenamefont {Gregg},\ and\ \citenamefont
  {Levi}}]{Li_Levi_2011_organic_measure_density_mobility}%
  \BibitemOpen
  \bibfield  {author} {\bibinfo {author} {\bibfnamefont {J.~V.}\ \bibnamefont
  {Li}}, \bibinfo {author} {\bibfnamefont {A.~M.}\ \bibnamefont {Nardes}},
  \bibinfo {author} {\bibfnamefont {Z.}~\bibnamefont {Liang}}, \bibinfo
  {author} {\bibfnamefont {S.~E.}\ \bibnamefont {Shaheen}}, \bibinfo {author}
  {\bibfnamefont {B.~A.}\ \bibnamefont {Gregg}},\ and\ \bibinfo {author}
  {\bibfnamefont {D.~H.}\ \bibnamefont {Levi}},\ }\bibfield  {title} {\bibinfo
  {title} {Simultaneous measurement of carrier density and mobility of organic
  semiconductors using capacitance techniques},\ }\href
  {https://doi.org/10.1016/j.orgel.2011.08.002} {\bibfield  {journal} {\bibinfo
   {journal} {Org. Electron.}\ }\textbf {\bibinfo {volume} {12}},\ \bibinfo
  {pages} {1879} (\bibinfo {year} {2011})}\BibitemShut {NoStop}%
\bibitem [{\citenamefont {Fukuzawa}\ \emph {et~al.}(2009)\citenamefont
  {Fukuzawa}, \citenamefont {Koshino},\ and\ \citenamefont
  {Ando}}]{Fukuzawa_Ando_2009_Hall_effect_graphene_SCBA}%
  \BibitemOpen
  \bibfield  {author} {\bibinfo {author} {\bibfnamefont {T.}~\bibnamefont
  {Fukuzawa}}, \bibinfo {author} {\bibfnamefont {M.}~\bibnamefont {Koshino}},\
  and\ \bibinfo {author} {\bibfnamefont {T.}~\bibnamefont {Ando}},\ }\bibfield
  {title} {\bibinfo {title} {Weak-field {Hall} effect in graphene calculated
  within self-consistent {Born} approximation},\ }\href
  {https://doi.org/10.1143/JPSJ.78.094714} {\bibfield  {journal} {\bibinfo
  {journal} {J. Phys. Soc. Jpn}\ }\textbf {\bibinfo {volume} {78}},\ \bibinfo
  {pages} {094714} (\bibinfo {year} {2009})}\BibitemShut {NoStop}%
\bibitem [{\citenamefont {Noro}\ and\ \citenamefont
  {Ando}(2016)}]{Noro_Ando_2016_Hall_long_range_scattering_graphene}%
  \BibitemOpen
  \bibfield  {author} {\bibinfo {author} {\bibfnamefont {M.}~\bibnamefont
  {Noro}}\ and\ \bibinfo {author} {\bibfnamefont {T.}~\bibnamefont {Ando}},\
  }\bibfield  {title} {\bibinfo {title} {Weak-field {Hall} effect in graphene
  with long-range scatterers},\ }\href {https://doi.org/10.7566/JPSJ.85.014708}
  {\bibfield  {journal} {\bibinfo  {journal} {J. Phys. Soc. Jpn}\ }\textbf
  {\bibinfo {volume} {85}},\ \bibinfo {pages} {014708} (\bibinfo {year}
  {2016})}\BibitemShut {NoStop}%
\bibitem [{\citenamefont {Morpurgo}\ \emph {et~al.}(2024)\citenamefont
  {Morpurgo}, \citenamefont {Rademaker}, \citenamefont {Berthod},\ and\
  \citenamefont {Giamarchi}}]{Morpurgo_Giamarchi_2024_Hall_effect_low_density}%
  \BibitemOpen
  \bibfield  {author} {\bibinfo {author} {\bibfnamefont {G.}~\bibnamefont
  {Morpurgo}}, \bibinfo {author} {\bibfnamefont {L.}~\bibnamefont {Rademaker}},
  \bibinfo {author} {\bibfnamefont {C.}~\bibnamefont {Berthod}},\ and\ \bibinfo
  {author} {\bibfnamefont {T.}~\bibnamefont {Giamarchi}},\ }\bibfield  {title}
  {\bibinfo {title} {{Hall} response of locally correlated two-dimensional
  electrons at low density},\ }\href
  {https://doi.org/10.1103/PhysRevResearch.6.013112} {\bibfield  {journal}
  {\bibinfo  {journal} {Phys. Rev. Research}\ }\textbf {\bibinfo {volume}
  {6}},\ \bibinfo {pages} {013112} (\bibinfo {year} {2024})}\BibitemShut
  {NoStop}%
\bibitem [{\citenamefont {Zhang}\ \emph {et~al.}(2019)\citenamefont {Zhang},
  \citenamefont {Berthod}, \citenamefont {Berger}, \citenamefont {Giamarchi},\
  and\ \citenamefont
  {Morpurgo}}]{Zhang_Morpurgo_2019_cross_quantum_capacitance}%
  \BibitemOpen
  \bibfield  {author} {\bibinfo {author} {\bibfnamefont {H.}~\bibnamefont
  {Zhang}}, \bibinfo {author} {\bibfnamefont {C.}~\bibnamefont {Berthod}},
  \bibinfo {author} {\bibfnamefont {H.}~\bibnamefont {Berger}}, \bibinfo
  {author} {\bibfnamefont {T.}~\bibnamefont {Giamarchi}},\ and\ \bibinfo
  {author} {\bibfnamefont {A.~F.}\ \bibnamefont {Morpurgo}},\ }\bibfield
  {title} {\bibinfo {title} {Band filling and cross quantum capacitance in
  ion-gated semiconducting transition metal dichalcogenide monolayers},\ }\href
  {https://doi.org/10.1021/acs.nanolett.9b03667} {\bibfield  {journal}
  {\bibinfo  {journal} {Nano Lett.}\ }\textbf {\bibinfo {volume} {19}},\
  \bibinfo {pages} {8836} (\bibinfo {year} {2019})}\BibitemShut {NoStop}%
\bibitem [{\citenamefont {Berthod}\ \emph {et~al.}(2021)\citenamefont
  {Berthod}, \citenamefont {Zhang}, \citenamefont {Morpurgo},\ and\
  \citenamefont
  {Giamarchi}}]{Berthod_Giamarchi_2021_cross_quantum_capacitance}%
  \BibitemOpen
  \bibfield  {author} {\bibinfo {author} {\bibfnamefont {C.}~\bibnamefont
  {Berthod}}, \bibinfo {author} {\bibfnamefont {H.}~\bibnamefont {Zhang}},
  \bibinfo {author} {\bibfnamefont {A.~F.}\ \bibnamefont {Morpurgo}},\ and\
  \bibinfo {author} {\bibfnamefont {T.}~\bibnamefont {Giamarchi}},\ }\bibfield
  {title} {\bibinfo {title} {Theory of cross quantum capacitance},\ }\href
  {https://doi.org/10.1103/PhysRevResearch.3.043036} {\bibfield  {journal}
  {\bibinfo  {journal} {Phys. Rev. Research}\ }\textbf {\bibinfo {volume}
  {3}},\ \bibinfo {pages} {043036} (\bibinfo {year} {2021})}\BibitemShut
  {NoStop}%
\bibitem [{\citenamefont {Choi}\ \emph {et~al.}(2018)\citenamefont {Choi},
  \citenamefont {Rodionov}, \citenamefont {Paterson}, \citenamefont {Panidi},
  \citenamefont {Saranin}, \citenamefont {Kharlamov}, \citenamefont {Didenko},
  \citenamefont {Anthopoulos}, \citenamefont {Cho},\ and\ \citenamefont
  {Podzorov}}]{Choi_Podzorov_2018_measure_charge_carrier_mobility_accurate}%
  \BibitemOpen
  \bibfield  {author} {\bibinfo {author} {\bibfnamefont {H.~H.}\ \bibnamefont
  {Choi}}, \bibinfo {author} {\bibfnamefont {Y.~I.}\ \bibnamefont {Rodionov}},
  \bibinfo {author} {\bibfnamefont {A.~F.}\ \bibnamefont {Paterson}}, \bibinfo
  {author} {\bibfnamefont {J.}~\bibnamefont {Panidi}}, \bibinfo {author}
  {\bibfnamefont {D.}~\bibnamefont {Saranin}}, \bibinfo {author} {\bibfnamefont
  {N.}~\bibnamefont {Kharlamov}}, \bibinfo {author} {\bibfnamefont {S.~I.}\
  \bibnamefont {Didenko}}, \bibinfo {author} {\bibfnamefont {T.~D.}\
  \bibnamefont {Anthopoulos}}, \bibinfo {author} {\bibfnamefont
  {K.}~\bibnamefont {Cho}},\ and\ \bibinfo {author} {\bibfnamefont
  {V.}~\bibnamefont {Podzorov}},\ }\bibfield  {title} {\bibinfo {title}
  {Accurate extraction of charge carrier mobility in 4-probe field-effect
  transistors},\ }\href {https://doi.org/10.1002/adfm.201707105} {\bibfield
  {journal} {\bibinfo  {journal} {Adv. Funct. Mater.}\ }\textbf {\bibinfo
  {volume} {28}},\ \bibinfo {pages} {1707105} (\bibinfo {year}
  {2018})}\BibitemShut {NoStop}%
\bibitem [{\citenamefont
  {Ziman}(2001)}]{Book_Ziman_2001_transport_electron_and_phonons}%
  \BibitemOpen
  \bibfield  {author} {\bibinfo {author} {\bibfnamefont {J.}~\bibnamefont
  {Ziman}},\ }\href {https://doi.org/10.1093/acprof:oso/9780198507796.001.0001}
  {\emph {\bibinfo {title} {{Electrons and Phonons: The Theory of Transport
  Phenomena in Solids}}}}\ (\bibinfo  {publisher} {Oxford University Press},\
  \bibinfo {address} {Oxford},\ \bibinfo {year} {2001})\BibitemShut {NoStop}%
\bibitem [{\citenamefont {Fukuyama}\ \emph {et~al.}(1969)\citenamefont
  {Fukuyama}, \citenamefont {Ebisawa},\ and\ \citenamefont
  {Wada}}]{Fukuyama_Wada_1969_Hall_effect_Kubo_I}%
  \BibitemOpen
  \bibfield  {author} {\bibinfo {author} {\bibfnamefont {H.}~\bibnamefont
  {Fukuyama}}, \bibinfo {author} {\bibfnamefont {H.}~\bibnamefont {Ebisawa}},\
  and\ \bibinfo {author} {\bibfnamefont {Y.}~\bibnamefont {Wada}},\ }\bibfield
  {title} {\bibinfo {title} {Theory of {Hall} effect. {I}: {N}early free
  electron},\ }\href {https://doi.org/10.1143/PTP.42.494} {\bibfield  {journal}
  {\bibinfo  {journal} {Prog. Theor. Phys.}\ }\textbf {\bibinfo {volume}
  {42}},\ \bibinfo {pages} {494} (\bibinfo {year} {1969})}\BibitemShut
  {NoStop}%
\bibitem [{\citenamefont {Fukuyama}(1969)}]{Fukuyama_1969_Hall_effect_Kubo_II}%
  \BibitemOpen
  \bibfield  {author} {\bibinfo {author} {\bibfnamefont {H.}~\bibnamefont
  {Fukuyama}},\ }\bibfield  {title} {\bibinfo {title} {Theory of {Hall} effect.
  {II}: {Bloch} electrons},\ }\href {https://doi.org/10.1143/PTP.42.1284}
  {\bibfield  {journal} {\bibinfo  {journal} {Prog. Theor. Phys.}\ }\textbf
  {\bibinfo {volume} {42}},\ \bibinfo {pages} {1284} (\bibinfo {year}
  {1969})}\BibitemShut {NoStop}%
\bibitem [{\citenamefont
  {Itoh}(1990)}]{Itoh_1990_Hall_effect_ladder_approximation}%
  \BibitemOpen
  \bibfield  {author} {\bibinfo {author} {\bibfnamefont {M.}~\bibnamefont
  {Itoh}},\ }\bibfield  {title} {\bibinfo {title} {The {Hall} effect in the
  ladder approximation},\ }\href {https://doi.org/10.1088/0953-8984/2/24/008}
  {\bibfield  {journal} {\bibinfo  {journal} {J. Phys-Condens. Mat.}\ }\textbf
  {\bibinfo {volume} {2}},\ \bibinfo {pages} {5357} (\bibinfo {year}
  {1990})}\BibitemShut {NoStop}%
\bibitem [{\citenamefont {Khodas}\ and\ \citenamefont
  {Finkel'stein}(2003)}]{Khodas_Finkelstein_2003_Hall_coefficient_ee_interactions}%
  \BibitemOpen
  \bibfield  {author} {\bibinfo {author} {\bibfnamefont {M.}~\bibnamefont
  {Khodas}}\ and\ \bibinfo {author} {\bibfnamefont {A.~M.}\ \bibnamefont
  {Finkel'stein}},\ }\bibfield  {title} {\bibinfo {title} {Hall coefficient in
  an interacting electron gas},\ }\href
  {https://doi.org/10.1103/PhysRevB.68.155114} {\bibfield  {journal} {\bibinfo
  {journal} {Phys. Rev. B}\ }\textbf {\bibinfo {volume} {68}},\ \bibinfo
  {pages} {155114} (\bibinfo {year} {2003})}\BibitemShut {NoStop}%
\bibitem [{\citenamefont {Kova\v{c}evi\'{c}}\ \emph {et~al.}(2025)\citenamefont
  {Kova\v{c}evi\'{c}}, \citenamefont {Ferrero},\ and\ \citenamefont
  {Vu\v{c}i\v{c}evi\'{c}}}]{Kovacevic_Vucicevic_2025_vertex_correction_Hubbard}%
  \BibitemOpen
  \bibfield  {author} {\bibinfo {author} {\bibfnamefont {J.}~\bibnamefont
  {Kova\v{c}evi\'{c}}}, \bibinfo {author} {\bibfnamefont {M.}~\bibnamefont
  {Ferrero}},\ and\ \bibinfo {author} {\bibfnamefont {J.}~\bibnamefont
  {Vu\v{c}i\v{c}evi\'{c}}},\ }\bibfield  {title} {\bibinfo {title} {Towards
  numerically exact computation of conductivity in the thermodynamic limit of
  interacting lattice models},\ }\href
  {https://doi.org/10.48550/arXiv.2501.19118} {\bibfield  {journal} {\bibinfo
  {journal} {arXiv:2501.19118}\ } (\bibinfo {year} {2025})}\BibitemShut
  {NoStop}%
\bibitem [{\citenamefont {Prelov\v{s}ek}\ and\ \citenamefont
  {Zotos}(2001)}]{Prelovsek_Zotos_2001_Hall_constant_correlated_systems}%
  \BibitemOpen
  \bibfield  {author} {\bibinfo {author} {\bibfnamefont {P.}~\bibnamefont
  {Prelov\v{s}ek}}\ and\ \bibinfo {author} {\bibfnamefont {X.}~\bibnamefont
  {Zotos}},\ }\bibfield  {title} {\bibinfo {title} {Reactive {Hall} constant of
  strongly correlated electrons},\ }\href
  {https://doi.org/10.1103/PhysRevB.64.235114} {\bibfield  {journal} {\bibinfo
  {journal} {Phys. Rev. B}\ }\textbf {\bibinfo {volume} {64}},\ \bibinfo
  {pages} {235114} (\bibinfo {year} {2001})}\BibitemShut {NoStop}%
\bibitem [{\citenamefont
  {Auerbach}(2018)}]{Auerbach_2018_Hall_number_metals_via_susceptibilities}%
  \BibitemOpen
  \bibfield  {author} {\bibinfo {author} {\bibfnamefont {A.}~\bibnamefont
  {Auerbach}},\ }\bibfield  {title} {\bibinfo {title} {{Hall} number of
  strongly correlated metals},\ }\href
  {https://doi.org/10.1103/PhysRevLett.121.066601} {\bibfield  {journal}
  {\bibinfo  {journal} {Phys. Rev. Lett.}\ }\textbf {\bibinfo {volume} {121}},\
  \bibinfo {pages} {066601} (\bibinfo {year} {2018})}\BibitemShut {NoStop}%
\bibitem [{\citenamefont {Castillo}\ and\ \citenamefont
  {Balseiro}(1992)}]{Castillo_Balseiro_1992_Hall_Hubbard_large_U}%
  \BibitemOpen
  \bibfield  {author} {\bibinfo {author} {\bibfnamefont {H.~E.}\ \bibnamefont
  {Castillo}}\ and\ \bibinfo {author} {\bibfnamefont {C.~A.}\ \bibnamefont
  {Balseiro}},\ }\bibfield  {title} {\bibinfo {title} {{Hall} conductivity and
  {Fermi} surface in highly correlated systems},\ }\href
  {https://doi.org/10.1103/PhysRevLett.68.121} {\bibfield  {journal} {\bibinfo
  {journal} {Phys. Rev. Lett.}\ }\textbf {\bibinfo {volume} {68}},\ \bibinfo
  {pages} {121} (\bibinfo {year} {1992})}\BibitemShut {NoStop}%
\bibitem [{\citenamefont {Assaad}\ and\ \citenamefont
  {Imada}(1995)}]{Assaad_Imada_1995_Hall_2D_Hubbard}%
  \BibitemOpen
  \bibfield  {author} {\bibinfo {author} {\bibfnamefont {F.~F.}\ \bibnamefont
  {Assaad}}\ and\ \bibinfo {author} {\bibfnamefont {M.}~\bibnamefont {Imada}},\
  }\bibfield  {title} {\bibinfo {title} {{Hall} coefficient for the
  two-dimensional {Hubbard} model},\ }\href
  {https://doi.org/10.1103/PhysRevLett.74.3868} {\bibfield  {journal} {\bibinfo
   {journal} {Phys. Rev. Lett.}\ }\textbf {\bibinfo {volume} {74}},\ \bibinfo
  {pages} {3868} (\bibinfo {year} {1995})}\BibitemShut {NoStop}%
\bibitem [{\citenamefont {Wang}\ \emph {et~al.}(2020)\citenamefont {Wang},
  \citenamefont {Ding}, \citenamefont {Moritz}, \citenamefont {Huang},\ and\
  \citenamefont {Devereaux}}]{Wang_Deveraux_2020_Hall_effect_Hubbard_QMC}%
  \BibitemOpen
  \bibfield  {author} {\bibinfo {author} {\bibfnamefont {W.~O.}\ \bibnamefont
  {Wang}}, \bibinfo {author} {\bibfnamefont {J.~K.}\ \bibnamefont {Ding}},
  \bibinfo {author} {\bibfnamefont {B.}~\bibnamefont {Moritz}}, \bibinfo
  {author} {\bibfnamefont {E.~W.}\ \bibnamefont {Huang}},\ and\ \bibinfo
  {author} {\bibfnamefont {T.~P.}\ \bibnamefont {Devereaux}},\ }\bibfield
  {title} {\bibinfo {title} {{DC} {Hall} coefficient of the strongly correlated
  {Hubbard} model},\ }\href {https://doi.org/10.1038/s41535-020-00254-w}
  {\bibfield  {journal} {\bibinfo  {journal} {npj Quantum Materials}\ }\textbf
  {\bibinfo {volume} {5}},\ \bibinfo {pages} {51} (\bibinfo {year}
  {2020})}\BibitemShut {NoStop}%
\bibitem [{\citenamefont {Pruschke}\ \emph {et~al.}(1995)\citenamefont
  {Pruschke}, \citenamefont {Jarrell},\ and\ \citenamefont
  {Freericks}}]{Pruschke_Freericks_1995_tranport_Hubbard_DMFT}%
  \BibitemOpen
  \bibfield  {author} {\bibinfo {author} {\bibfnamefont {T.}~\bibnamefont
  {Pruschke}}, \bibinfo {author} {\bibfnamefont {M.}~\bibnamefont {Jarrell}},\
  and\ \bibinfo {author} {\bibfnamefont {J.}~\bibnamefont {Freericks}},\
  }\bibfield  {title} {\bibinfo {title} {Anomalous normal-state properties of
  high-{$T_c$} superconductors: intrinsic properties of strongly correlated
  electron systems?},\ }\href {https://doi.org/10.1080/00018739500101526}
  {\bibfield  {journal} {\bibinfo  {journal} {Adv. Phys.}\ }\textbf {\bibinfo
  {volume} {44}},\ \bibinfo {pages} {187} (\bibinfo {year} {1995})}\BibitemShut
  {NoStop}%
\bibitem [{\citenamefont {Markov}\ \emph {et~al.}(2019)\citenamefont {Markov},
  \citenamefont {Rohringer},\ and\ \citenamefont
  {Rubtsov}}]{Markov_Rubstov_2019_Hofstadter_butterfly_Hubbard_DMFT}%
  \BibitemOpen
  \bibfield  {author} {\bibinfo {author} {\bibfnamefont {A.~A.}\ \bibnamefont
  {Markov}}, \bibinfo {author} {\bibfnamefont {G.}~\bibnamefont {Rohringer}},\
  and\ \bibinfo {author} {\bibfnamefont {A.~N.}\ \bibnamefont {Rubtsov}},\
  }\bibfield  {title} {\bibinfo {title} {Robustness of the topological
  quantization of the {Hall} conductivity for correlated lattice electrons at
  finite temperatures},\ }\href {https://doi.org/10.1103/PhysRevB.100.115102}
  {\bibfield  {journal} {\bibinfo  {journal} {Phys. Rev. B}\ }\textbf {\bibinfo
  {volume} {100}},\ \bibinfo {pages} {115102} (\bibinfo {year}
  {2019})}\BibitemShut {NoStop}%
\bibitem [{\citenamefont {Vu\v{c}i\v{c}evi\'{c}}\ and\ \citenamefont
  {\v{Z}itko}(2021)}]{Vucicevic_Zitko_2021_DMFT_conductivity_Hubbard}%
  \BibitemOpen
  \bibfield  {author} {\bibinfo {author} {\bibfnamefont {J.}~\bibnamefont
  {Vu\v{c}i\v{c}evi\'{c}}}\ and\ \bibinfo {author} {\bibfnamefont
  {R.}~\bibnamefont {\v{Z}itko}},\ }\bibfield  {title} {\bibinfo {title}
  {Electrical conductivity in the {Hubbard} model: {Orbital} effects of
  magnetic field},\ }\href {https://doi.org/10.1103/PhysRevB.104.205101}
  {\bibfield  {journal} {\bibinfo  {journal} {Phys. Rev. B}\ }\textbf {\bibinfo
  {volume} {104}},\ \bibinfo {pages} {205101} (\bibinfo {year}
  {2021})}\BibitemShut {NoStop}%
\bibitem [{\citenamefont {Manning}\ and\ \citenamefont
  {Chen}(1993)}]{Manning_Chen_1993_Hall_effect_t_J_model}%
  \BibitemOpen
  \bibfield  {author} {\bibinfo {author} {\bibfnamefont {S.~M.}\ \bibnamefont
  {Manning}}\ and\ \bibinfo {author} {\bibfnamefont {Y.}~\bibnamefont {Chen}},\
  }\bibfield  {title} {\bibinfo {title} {The {Hall} effect in the normal state
  of the {$t$--$J$} model},\ }\href {https://doi.org/10.1088/0953-8984/5/3/001}
  {\bibfield  {journal} {\bibinfo  {journal} {J. Phys-Condens. Mat.}\ }\textbf
  {\bibinfo {volume} {5}},\ \bibinfo {pages} {L23} (\bibinfo {year}
  {1993})}\BibitemShut {NoStop}%
\bibitem [{\citenamefont {Kontani}\ \emph {et~al.}(1999)\citenamefont
  {Kontani}, \citenamefont {Kanki},\ and\ \citenamefont
  {Ueda}}]{Kontani_Ueda_1999_Hall_effect_high_Tc}%
  \BibitemOpen
  \bibfield  {author} {\bibinfo {author} {\bibfnamefont {H.}~\bibnamefont
  {Kontani}}, \bibinfo {author} {\bibfnamefont {K.}~\bibnamefont {Kanki}},\
  and\ \bibinfo {author} {\bibfnamefont {K.}~\bibnamefont {Ueda}},\ }\bibfield
  {title} {\bibinfo {title} {{Hall} effect and resistivity in high-{$T_c$}
  superconductors: The conserving approximation},\ }\href
  {https://doi.org/10.1103/PhysRevB.59.14723} {\bibfield  {journal} {\bibinfo
  {journal} {Phys. Rev. B}\ }\textbf {\bibinfo {volume} {59}},\ \bibinfo
  {pages} {14723} (\bibinfo {year} {1999})}\BibitemShut {NoStop}%
\bibitem [{\citenamefont
  {Kontani}(2007)}]{Kontani_2007_Hall_effect_high_Tc_FLEX}%
  \BibitemOpen
  \bibfield  {author} {\bibinfo {author} {\bibfnamefont {H.}~\bibnamefont
  {Kontani}},\ }\bibfield  {title} {\bibinfo {title} {Theory of infrared {Hall}
  conductivity based on the {Fermi} liquid theory: Analysis of high-{$T_c$}
  superconductors},\ }\href {https://doi.org/10.1143/JPSJ.76.074707} {\bibfield
   {journal} {\bibinfo  {journal} {J. Phys. Soc. Jpn}\ }\textbf {\bibinfo
  {volume} {76}},\ \bibinfo {pages} {074707} (\bibinfo {year}
  {2007})}\BibitemShut {NoStop}%
\bibitem [{\citenamefont {Mitscherling}\ and\ \citenamefont
  {Metzner}(2018)}]{Mitscherling_Metzner_2018}%
  \BibitemOpen
  \bibfield  {author} {\bibinfo {author} {\bibfnamefont {J.}~\bibnamefont
  {Mitscherling}}\ and\ \bibinfo {author} {\bibfnamefont {W.}~\bibnamefont
  {Metzner}},\ }\bibfield  {title} {\bibinfo {title} {Longitudinal conductivity
  and {Hall} coefficient in two-dimensional metals with spiral magnetic
  order},\ }\href {https://doi.org/10.1103/PhysRevB.98.195126} {\bibfield
  {journal} {\bibinfo  {journal} {Phys. Rev. B}\ }\textbf {\bibinfo {volume}
  {98}},\ \bibinfo {pages} {195126} (\bibinfo {year} {2018})}\BibitemShut
  {NoStop}%
\bibitem [{\citenamefont {Le{\'o}n}\ and\ \citenamefont
  {Giamarchi}(2006)}]{Leon_Giamarchi_2006_Hall_quasi_1d_organic}%
  \BibitemOpen
  \bibfield  {author} {\bibinfo {author} {\bibfnamefont {G.}~\bibnamefont
  {Le{\'o}n}}\ and\ \bibinfo {author} {\bibfnamefont {T.}~\bibnamefont
  {Giamarchi}},\ }\bibfield  {title} {\bibinfo {title} {Hall effect in quasi
  one-dimensional organic conductors},\ }\href
  {https://doi.org/10.1007/BF02679514} {\bibfield  {journal} {\bibinfo
  {journal} {J. Phys. F}\ }\textbf {\bibinfo {volume} {142}},\ \bibinfo {pages}
  {315} (\bibinfo {year} {2006})}\BibitemShut {NoStop}%
\bibitem [{\citenamefont {Le\'on}\ \emph {et~al.}(2007)\citenamefont {Le\'on},
  \citenamefont {Berthod},\ and\ \citenamefont
  {Giamarchi}}]{Leon_Giamarchi_2007_Hall_strong_correlated_low_d_systems}%
  \BibitemOpen
  \bibfield  {author} {\bibinfo {author} {\bibfnamefont {G.}~\bibnamefont
  {Le\'on}}, \bibinfo {author} {\bibfnamefont {C.}~\bibnamefont {Berthod}},\
  and\ \bibinfo {author} {\bibfnamefont {T.}~\bibnamefont {Giamarchi}},\
  }\bibfield  {title} {\bibinfo {title} {Hall effect in strongly correlated
  low-dimensional systems},\ }\href
  {https://doi.org/10.1103/PhysRevB.75.195123} {\bibfield  {journal} {\bibinfo
  {journal} {Phys. Rev. B}\ }\textbf {\bibinfo {volume} {75}},\ \bibinfo
  {pages} {195123} (\bibinfo {year} {2007})}\BibitemShut {NoStop}%
\bibitem [{\citenamefont {Le\'on}\ \emph {et~al.}(2008)\citenamefont {Le\'on},
  \citenamefont {Berthod}, \citenamefont {Giamarchi},\ and\ \citenamefont
  {Millis}}]{Leon_Millis_2008_Hall_triangular_lattice}%
  \BibitemOpen
  \bibfield  {author} {\bibinfo {author} {\bibfnamefont {G.}~\bibnamefont
  {Le\'on}}, \bibinfo {author} {\bibfnamefont {C.}~\bibnamefont {Berthod}},
  \bibinfo {author} {\bibfnamefont {T.}~\bibnamefont {Giamarchi}},\ and\
  \bibinfo {author} {\bibfnamefont {A.~J.}\ \bibnamefont {Millis}},\ }\bibfield
   {title} {\bibinfo {title} {Hall effect on the triangular lattice},\ }\href
  {https://doi.org/10.1103/PhysRevB.78.085105} {\bibfield  {journal} {\bibinfo
  {journal} {Phys. Rev. B}\ }\textbf {\bibinfo {volume} {78}},\ \bibinfo
  {pages} {085105} (\bibinfo {year} {2008})}\BibitemShut {NoStop}%
\bibitem [{\citenamefont {Filippone}\ \emph {et~al.}(2019)\citenamefont
  {Filippone}, \citenamefont {Bardyn}, \citenamefont {Greschner},\ and\
  \citenamefont {Giamarchi}}]{Filippone_Giamarchi_2019_vanishing_Hall}%
  \BibitemOpen
  \bibfield  {author} {\bibinfo {author} {\bibfnamefont {M.}~\bibnamefont
  {Filippone}}, \bibinfo {author} {\bibfnamefont {C.-E.}\ \bibnamefont
  {Bardyn}}, \bibinfo {author} {\bibfnamefont {S.}~\bibnamefont {Greschner}},\
  and\ \bibinfo {author} {\bibfnamefont {T.}~\bibnamefont {Giamarchi}},\
  }\bibfield  {title} {\bibinfo {title} {Vanishing {Hall} response of charged
  fermions in a transverse magnetic field},\ }\href
  {https://doi.org/10.1103/PhysRevLett.123.086803} {\bibfield  {journal}
  {\bibinfo  {journal} {Phys. Rev. Lett.}\ }\textbf {\bibinfo {volume} {123}},\
  \bibinfo {pages} {086803} (\bibinfo {year} {2019})}\BibitemShut {NoStop}%
\bibitem [{\citenamefont {Greschner}\ \emph {et~al.}(2019)\citenamefont
  {Greschner}, \citenamefont {Filippone},\ and\ \citenamefont
  {Giamarchi}}]{Greschner_Giamarchi_2019_Universal_Hall_response}%
  \BibitemOpen
  \bibfield  {author} {\bibinfo {author} {\bibfnamefont {S.}~\bibnamefont
  {Greschner}}, \bibinfo {author} {\bibfnamefont {M.}~\bibnamefont
  {Filippone}},\ and\ \bibinfo {author} {\bibfnamefont {T.}~\bibnamefont
  {Giamarchi}},\ }\bibfield  {title} {\bibinfo {title} {Universal {Hall}
  response in interacting quantum systems},\ }\href
  {https://doi.org/10.1103/PhysRevLett.122.083402} {\bibfield  {journal}
  {\bibinfo  {journal} {Phys. Rev. Lett.}\ }\textbf {\bibinfo {volume} {122}},\
  \bibinfo {pages} {083402} (\bibinfo {year} {2019})}\BibitemShut {NoStop}%
\bibitem [{\citenamefont {Buser}\ \emph {et~al.}(2021)\citenamefont {Buser},
  \citenamefont {Greschner}, \citenamefont {Schollw\"ock},\ and\ \citenamefont
  {Giamarchi}}]{Buser_Giamarchi_2021_Hall_voltage_synthetic_quantum_systems}%
  \BibitemOpen
  \bibfield  {author} {\bibinfo {author} {\bibfnamefont {M.}~\bibnamefont
  {Buser}}, \bibinfo {author} {\bibfnamefont {S.}~\bibnamefont {Greschner}},
  \bibinfo {author} {\bibfnamefont {U.}~\bibnamefont {Schollw\"ock}},\ and\
  \bibinfo {author} {\bibfnamefont {T.}~\bibnamefont {Giamarchi}},\ }\bibfield
  {title} {\bibinfo {title} {Probing the {Hall} voltage in synthetic quantum
  systems},\ }\href {https://doi.org/10.1103/PhysRevLett.126.030501} {\bibfield
   {journal} {\bibinfo  {journal} {Phys. Rev. Lett.}\ }\textbf {\bibinfo
  {volume} {126}},\ \bibinfo {pages} {030501} (\bibinfo {year}
  {2021})}\BibitemShut {NoStop}%
\bibitem [{\citenamefont
  {Khurana}(1990)}]{Khurana_1990_conductivity_infinite_d_Hubbard}%
  \BibitemOpen
  \bibfield  {author} {\bibinfo {author} {\bibfnamefont {A.}~\bibnamefont
  {Khurana}},\ }\bibfield  {title} {\bibinfo {title} {Electrical conductivity
  in the infinite-dimensional {Hubbard} model},\ }\href
  {https://doi.org/10.1103/PhysRevLett.64.1990} {\bibfield  {journal} {\bibinfo
   {journal} {Phys. Rev. Lett.}\ }\textbf {\bibinfo {volume} {64}},\ \bibinfo
  {pages} {1990} (\bibinfo {year} {1990})}\BibitemShut {NoStop}%
\bibitem [{\citenamefont {Liu}(1970)}]{Liu_1970_Kubo_conductivity_impurity}%
  \BibitemOpen
  \bibfield  {author} {\bibinfo {author} {\bibfnamefont {S.}~\bibnamefont
  {Liu}},\ }\bibfield  {title} {\bibinfo {title} {Nonlocal conductivities and
  dielectric functions of free electron gas with impurity scattering},\ }\href
  {https://doi.org/10.1016/0003-4916(70)90399-4} {\bibfield  {journal}
  {\bibinfo  {journal} {Ann. Phys. (NY)}\ }\textbf {\bibinfo {volume} {59}},\
  \bibinfo {pages} {165} (\bibinfo {year} {1970})}\BibitemShut {NoStop}%
\bibitem [{\citenamefont
  {Fukuyama}(1980)}]{Fukuyama_1980_Hall_effect_2d_disordered}%
  \BibitemOpen
  \bibfield  {author} {\bibinfo {author} {\bibfnamefont {H.}~\bibnamefont
  {Fukuyama}},\ }\bibfield  {title} {\bibinfo {title} {Hall effect in
  two-dimensional disordered systems},\ }\href
  {https://doi.org/10.1143/JPSJ.49.644} {\bibfield  {journal} {\bibinfo
  {journal} {J. Phys. Soc. Jpn}\ }\textbf {\bibinfo {volume} {49}},\ \bibinfo
  {pages} {644} (\bibinfo {year} {1980})}\BibitemShut {NoStop}%
\bibitem [{\citenamefont
  {Ando}(2014)}]{Ando_2014_Hall_long_range_scattering_bilayer_graphene}%
  \BibitemOpen
  \bibfield  {author} {\bibinfo {author} {\bibfnamefont {T.}~\bibnamefont
  {Ando}},\ }\bibfield  {title} {\bibinfo {title} {Bilayer graphene with
  long-range scatterers: Diamagnetism and weak-field {Hall} effect},\ }\href
  {https://doi.org/10.1016/j.physe.2013.11.015} {\bibfield  {journal} {\bibinfo
   {journal} {Physica E}\ }\textbf {\bibinfo {volume} {58}},\ \bibinfo {pages}
  {6} (\bibinfo {year} {2014})}\BibitemShut {NoStop}%
\bibitem [{\citenamefont {Pickem}\ \emph {et~al.}(2022)\citenamefont {Pickem},
  \citenamefont {Maggio},\ and\ \citenamefont {Tomczak}}]{Pickem_Tomczak_2022}%
  \BibitemOpen
  \bibfield  {author} {\bibinfo {author} {\bibfnamefont {M.}~\bibnamefont
  {Pickem}}, \bibinfo {author} {\bibfnamefont {E.}~\bibnamefont {Maggio}},\
  and\ \bibinfo {author} {\bibfnamefont {J.~M.}\ \bibnamefont {Tomczak}},\
  }\bibfield  {title} {\bibinfo {title} {Prototypical many-body signatures in
  transport properties of semiconductors},\ }\href
  {https://doi.org/10.1103/PhysRevB.105.085139} {\bibfield  {journal} {\bibinfo
   {journal} {Phys. Rev. B}\ }\textbf {\bibinfo {volume} {105}},\ \bibinfo
  {pages} {085139} (\bibinfo {year} {2022})}\BibitemShut {NoStop}%
\bibitem [{\citenamefont {Bruus}\ and\ \citenamefont
  {Flensberg}(2016)}]{Book_Bruus_Flensberg_2016_Many_Body}%
  \BibitemOpen
  \bibfield  {author} {\bibinfo {author} {\bibfnamefont {H.}~\bibnamefont
  {Bruus}}\ and\ \bibinfo {author} {\bibfnamefont {K.}~\bibnamefont
  {Flensberg}},\ }\href {https://doi.org/10.1093/oso/9780198566335.001.0001}
  {\emph {\bibinfo {title} {Many-Body Quantum Theory in Condensed Matter
  Physics : An Introduction}}}\ (\bibinfo  {publisher} {Oxford University
  Press},\ \bibinfo {address} {Oxford},\ \bibinfo {year} {2016})\BibitemShut
  {NoStop}%
\bibitem [{\citenamefont
  {Berthod}(2018)}]{Book_Berthod_2018_Spectroscopic_Probes}%
  \BibitemOpen
  \bibfield  {author} {\bibinfo {author} {\bibfnamefont {C.}~\bibnamefont
  {Berthod}},\ }\href {https://doi.org/10.1088/978-0-7503-1741-2} {\emph
  {\bibinfo {title} {Spectroscopic Probes of Quantum Matter}}}\ (\bibinfo
  {publisher} {IOP},\ \bibinfo {address} {Bristol, UK},\ \bibinfo {year}
  {2018})\BibitemShut {NoStop}%
\bibitem [{\citenamefont
  {Berthod}(2025{\natexlab{a}})}]{Berthod_2025_Piecewise}%
  \BibitemOpen
  \bibfield  {author} {\bibinfo {author} {\bibfnamefont {C.}~\bibnamefont
  {Berthod}},\ }\href@noop {} {\bibinfo {title} {Piecewise: {F}lexible
  piecewise functions for fast integral transforms in {Julia}}},\ \bibinfo
  {howpublished} {\url{https://github.com/ChristopheBerthod/Piecewise.jl}}
  (\bibinfo {year} {2025}{\natexlab{a}})\BibitemShut {NoStop}%
\bibitem [{\citenamefont
  {Berthod}(2025{\natexlab{b}})}]{Berthod_2025_Magnetotransport_piecewise}%
  \BibitemOpen
  \bibfield  {author} {\bibinfo {author} {\bibfnamefont {C.}~\bibnamefont
  {Berthod}},\ }\href@noop {} {\bibinfo {title} {Linear magneto-transport with
  a local self-energy in {Julia}}},\ \bibinfo {howpublished}
  {\url{https://github.com/ChristopheBerthod/MagnetoTransport.jl}} (\bibinfo
  {year} {2025}{\natexlab{b}})\BibitemShut {NoStop}%
\bibitem [{Note1()}]{Note1}%
  \BibitemOpen
  \bibinfo {note} {As $-\protect \mathrm {Im}\protect \,\Sigma (\varepsilon )$
  must remain strictly positive for causality reasons, an infinitesimal shift
  $i0$ must in principle be inserted into Eq.~(\ref {eq:SCBA}), such that
  $\protect \mathrm {Im}\protect \,\Sigma ^{\protect \mathrm
  {SCBA}}(\varepsilon )$ is never strictly zero.}\BibitemShut {Stop}%
\bibitem [{\citenamefont {Yamagishi}\ \emph {et~al.}(2010)\citenamefont
  {Yamagishi}, \citenamefont {Soeda}, \citenamefont {Uemura}, \citenamefont
  {Okada}, \citenamefont {Takatsuki}, \citenamefont {Nishikawa}, \citenamefont
  {Nakazawa}, \citenamefont {Doi}, \citenamefont {Takimiya},\ and\
  \citenamefont {Takeya}}]{Yamagishi_Takeya_2010_Hall_organic_thin_film}%
  \BibitemOpen
  \bibfield  {author} {\bibinfo {author} {\bibfnamefont {M.}~\bibnamefont
  {Yamagishi}}, \bibinfo {author} {\bibfnamefont {J.}~\bibnamefont {Soeda}},
  \bibinfo {author} {\bibfnamefont {T.}~\bibnamefont {Uemura}}, \bibinfo
  {author} {\bibfnamefont {Y.}~\bibnamefont {Okada}}, \bibinfo {author}
  {\bibfnamefont {Y.}~\bibnamefont {Takatsuki}}, \bibinfo {author}
  {\bibfnamefont {T.}~\bibnamefont {Nishikawa}}, \bibinfo {author}
  {\bibfnamefont {Y.}~\bibnamefont {Nakazawa}}, \bibinfo {author}
  {\bibfnamefont {I.}~\bibnamefont {Doi}}, \bibinfo {author} {\bibfnamefont
  {K.}~\bibnamefont {Takimiya}},\ and\ \bibinfo {author} {\bibfnamefont
  {J.}~\bibnamefont {Takeya}},\ }\bibfield  {title} {\bibinfo {title}
  {Free-electron-like {Hall} effect in high-mobility organic thin-film
  transistors},\ }\href {https://doi.org/10.1103/PhysRevB.81.161306} {\bibfield
   {journal} {\bibinfo  {journal} {Phys. Rev. B}\ }\textbf {\bibinfo {volume}
  {81}},\ \bibinfo {pages} {161306} (\bibinfo {year} {2010})}\BibitemShut
  {NoStop}%
\bibitem [{\citenamefont {Uemura}\ \emph {et~al.}(2012)\citenamefont {Uemura},
  \citenamefont {Yamagishi}, \citenamefont {Soeda}, \citenamefont {Takatsuki},
  \citenamefont {Okada}, \citenamefont {Nakazawa},\ and\ \citenamefont
  {Takeya}}]{Uemura_Takeya_2012_Temperature_dependance_Hall}%
  \BibitemOpen
  \bibfield  {author} {\bibinfo {author} {\bibfnamefont {T.}~\bibnamefont
  {Uemura}}, \bibinfo {author} {\bibfnamefont {M.}~\bibnamefont {Yamagishi}},
  \bibinfo {author} {\bibfnamefont {J.}~\bibnamefont {Soeda}}, \bibinfo
  {author} {\bibfnamefont {Y.}~\bibnamefont {Takatsuki}}, \bibinfo {author}
  {\bibfnamefont {Y.}~\bibnamefont {Okada}}, \bibinfo {author} {\bibfnamefont
  {Y.}~\bibnamefont {Nakazawa}},\ and\ \bibinfo {author} {\bibfnamefont
  {J.}~\bibnamefont {Takeya}},\ }\bibfield  {title} {\bibinfo {title}
  {Temperature dependence of the {Hall} effect in pentacene field-effect
  transistors: Possibility of charge decoherence induced by molecular
  fluctuations},\ }\href {https://doi.org/10.1103/PhysRevB.85.035313}
  {\bibfield  {journal} {\bibinfo  {journal} {Phys. Rev. B}\ }\textbf {\bibinfo
  {volume} {85}},\ \bibinfo {pages} {035313} (\bibinfo {year}
  {2012})}\BibitemShut {NoStop}%
\bibitem [{\citenamefont {Pradhan}\ \emph {et~al.}(2015)\citenamefont
  {Pradhan}, \citenamefont {Rhodes}, \citenamefont {Memaran}, \citenamefont
  {Poumirol}, \citenamefont {Smirnov}, \citenamefont {Talapatra}, \citenamefont
  {Feng}, \citenamefont {Perea-Lopez}, \citenamefont {Elias}, \citenamefont
  {Terrones}, \citenamefont {Ajayan},\ and\ \citenamefont
  {Balicas}}]{Pradhan_Balicas_2015_weak_field_Hall_TMD}%
  \BibitemOpen
  \bibfield  {author} {\bibinfo {author} {\bibfnamefont {N.~R.}\ \bibnamefont
  {Pradhan}}, \bibinfo {author} {\bibfnamefont {D.}~\bibnamefont {Rhodes}},
  \bibinfo {author} {\bibfnamefont {S.}~\bibnamefont {Memaran}}, \bibinfo
  {author} {\bibfnamefont {J.~M.}\ \bibnamefont {Poumirol}}, \bibinfo {author}
  {\bibfnamefont {D.}~\bibnamefont {Smirnov}}, \bibinfo {author} {\bibfnamefont
  {S.}~\bibnamefont {Talapatra}}, \bibinfo {author} {\bibfnamefont
  {S.}~\bibnamefont {Feng}}, \bibinfo {author} {\bibfnamefont {N.}~\bibnamefont
  {Perea-Lopez}}, \bibinfo {author} {\bibfnamefont {A.~L.}\ \bibnamefont
  {Elias}}, \bibinfo {author} {\bibfnamefont {M.}~\bibnamefont {Terrones}},
  \bibinfo {author} {\bibfnamefont {P.~M.}\ \bibnamefont {Ajayan}},\ and\
  \bibinfo {author} {\bibfnamefont {L.}~\bibnamefont {Balicas}},\ }\bibfield
  {title} {\bibinfo {title} {{Hall} and field-effect mobilities in few layered
  {$p$-WSe$_2$} field-effect transistors},\ }\href
  {https://doi.org/10.1038/srep08979} {\bibfield  {journal} {\bibinfo
  {journal} {Sci. Rep. UK}\ }\textbf {\bibinfo {volume} {5}},\ \bibinfo {pages}
  {8979} (\bibinfo {year} {2015})}\BibitemShut {NoStop}%
\bibitem [{\citenamefont {Masseroni}\ \emph {et~al.}(2024)\citenamefont
  {Masseroni}, \citenamefont {Gull}, \citenamefont {Panigrahi}, \citenamefont
  {Jacobsen}, \citenamefont {Fischer}, \citenamefont {Tong}, \citenamefont
  {Gerber}, \citenamefont {Niese}, \citenamefont {Taniguchi}, \citenamefont
  {Watanabe}, \citenamefont {Levitov}, \citenamefont {Ihn}, \citenamefont
  {Ensslin},\ and\ \citenamefont
  {Duprez}}]{Masseroni_Duprez_2024_TMD_graphene_Spin_Orbit}%
  \BibitemOpen
  \bibfield  {author} {\bibinfo {author} {\bibfnamefont {M.}~\bibnamefont
  {Masseroni}}, \bibinfo {author} {\bibfnamefont {M.}~\bibnamefont {Gull}},
  \bibinfo {author} {\bibfnamefont {A.}~\bibnamefont {Panigrahi}}, \bibinfo
  {author} {\bibfnamefont {N.}~\bibnamefont {Jacobsen}}, \bibinfo {author}
  {\bibfnamefont {F.}~\bibnamefont {Fischer}}, \bibinfo {author} {\bibfnamefont
  {C.}~\bibnamefont {Tong}}, \bibinfo {author} {\bibfnamefont {J.~D.}\
  \bibnamefont {Gerber}}, \bibinfo {author} {\bibfnamefont {M.}~\bibnamefont
  {Niese}}, \bibinfo {author} {\bibfnamefont {T.}~\bibnamefont {Taniguchi}},
  \bibinfo {author} {\bibfnamefont {K.}~\bibnamefont {Watanabe}}, \bibinfo
  {author} {\bibfnamefont {L.}~\bibnamefont {Levitov}}, \bibinfo {author}
  {\bibfnamefont {T.}~\bibnamefont {Ihn}}, \bibinfo {author} {\bibfnamefont
  {K.}~\bibnamefont {Ensslin}},\ and\ \bibinfo {author} {\bibfnamefont
  {H.}~\bibnamefont {Duprez}},\ }\bibfield  {title} {\bibinfo {title}
  {Spin-orbit proximity in {MoS$_2$}/bilayer graphene heterostructures},\
  }\href {https://doi.org/10.1038/s41467-024-53324-z} {\bibfield  {journal}
  {\bibinfo  {journal} {Nature Communications}\ }\textbf {\bibinfo {volume}
  {15}},\ \bibinfo {pages} {9251} (\bibinfo {year} {2024})}\BibitemShut
  {NoStop}%
\bibitem [{\citenamefont {Guti{\'e}rrez-Lezama}\ \emph
  {et~al.}(2021)\citenamefont {Guti{\'e}rrez-Lezama}, \citenamefont {Ubrig},
  \citenamefont {Ponomarev},\ and\ \citenamefont
  {Morpurgo}}]{Review_Gutierrez_Morpurgo_2021_technique_measure_band_gaps}%
  \BibitemOpen
  \bibfield  {author} {\bibinfo {author} {\bibfnamefont {I.}~\bibnamefont
  {Guti{\'e}rrez-Lezama}}, \bibinfo {author} {\bibfnamefont {N.}~\bibnamefont
  {Ubrig}}, \bibinfo {author} {\bibfnamefont {E.}~\bibnamefont {Ponomarev}},\
  and\ \bibinfo {author} {\bibfnamefont {A.~F.}\ \bibnamefont {Morpurgo}},\
  }\bibfield  {title} {\bibinfo {title} {Ionic gate spectroscopy of {2D}
  semiconductors},\ }\href {https://doi.org/10.1038/s42254-021-00317-2}
  {\bibfield  {journal} {\bibinfo  {journal} {Nat. Rev. Phys.}\ }\textbf
  {\bibinfo {volume} {3}},\ \bibinfo {pages} {508} (\bibinfo {year}
  {2021})}\BibitemShut {NoStop}%
\bibitem [{\citenamefont {Anderson}(1958)}]{Anderson_1958_localisation}%
  \BibitemOpen
  \bibfield  {author} {\bibinfo {author} {\bibfnamefont {P.~W.}\ \bibnamefont
  {Anderson}},\ }\bibfield  {title} {\bibinfo {title} {Absence of diffusion in
  certain random lattices},\ }\href {https://doi.org/10.1103/PhysRev.109.1492}
  {\bibfield  {journal} {\bibinfo  {journal} {Phys. Rev.}\ }\textbf {\bibinfo
  {volume} {109}},\ \bibinfo {pages} {1492} (\bibinfo {year}
  {1958})}\BibitemShut {NoStop}%
\bibitem [{\citenamefont {Lee}\ and\ \citenamefont
  {Ramakrishnan}(1985)}]{Review_Lee_Ramakrishnan_1985_disorder}%
  \BibitemOpen
  \bibfield  {author} {\bibinfo {author} {\bibfnamefont {P.~A.}\ \bibnamefont
  {Lee}}\ and\ \bibinfo {author} {\bibfnamefont {T.~V.}\ \bibnamefont
  {Ramakrishnan}},\ }\bibfield  {title} {\bibinfo {title} {Disordered
  electronic systems},\ }\href {https://doi.org/10.1103/RevModPhys.57.287}
  {\bibfield  {journal} {\bibinfo  {journal} {Rev. Mod. Phys.}\ }\textbf
  {\bibinfo {volume} {57}},\ \bibinfo {pages} {287} (\bibinfo {year}
  {1985})}\BibitemShut {NoStop}%
\bibitem [{\citenamefont {Morpurgo}\ \emph {et~al.}(2025)\citenamefont
  {Morpurgo}, \citenamefont {Berthod},\ and\ \citenamefont
  {Giamarchi}}]{yareta}%
  \BibitemOpen
  \bibfield  {author} {\bibinfo {author} {\bibfnamefont {G.}~\bibnamefont
  {Morpurgo}}, \bibinfo {author} {\bibfnamefont {C.}~\bibnamefont {Berthod}},\
  and\ \bibinfo {author} {\bibfnamefont {T.}~\bibnamefont {Giamarchi}},\
  }\bibfield  {title} {\bibinfo {title} {Open data to ``{U}niversal low-density
  power laws of the dc conductivity and {Hall} constant in the self-consistent
  {Born} approximation''},\ }\bibfield  {journal} {\bibinfo  {journal} {Yareta,
  {University} of {Geneva}}\ }\href
  {https://doi.org/10.26037/yareta:gl44nnfso5enhohdbfruvfnvge}
  {10.26037/yareta:gl44nnfso5enhohdbfruvfnvge} (\bibinfo {year}
  {2025})\BibitemShut {NoStop}%
\end{thebibliography}

%

\end{document}